\documentclass[a4paper,11pt,spacing]{article}
\pdfoutput=1 % if your are submitting a pdflatex (i.e. if you have
\usepackage{jcappub} 
\usepackage[T1]{fontenc}

\usepackage{natbib}
\usepackage{cancel}
\usepackage{enumitem}    
\usepackage{amsthm}         
\usepackage{amsmath} 	   
\usepackage{amssymb}       
\usepackage{amsfonts}
\usepackage{graphicx} 	    
\usepackage{verbatim}
\usepackage{color}
\usepackage{colortbl}
\usepackage{float}
\usepackage{multirow}

\usepackage{setspace}
\usepackage{subcaption}

\definecolor{babyblue}{rgb}{0.54, 0.81, 0.94}
\definecolor{corn}{rgb}{0.98, 0.93, 0.36}

\title{The anamorphic universe}

\author[a,1]{Anna Ijjas,}
\note{Corresponding author.}
\author[a,b]{Paul J. Steinhardt}

\affiliation[a]{Princeton Center for Theoretical Science, Princeton University\\Princeton, NJ, 08544 USA}
\affiliation[b]{Department of Physics, Princeton University\\Princeton, NJ, 08544 USA}

\emailAdd{aijjas@princeton.edu}
\emailAdd{steinh@princeton.edu}

\abstract{
We introduce ``anamorphic'' cosmology, an approach for explaining the smoothness and flatness of the universe on large scales and the generation of a nearly scale-invariant spectrum of adiabatic density perturbations.  The defining feature is a smoothing phase that acts like a contracting universe based on  some Weyl frame-invariant criteria and an expanding universe based on other frame-invariant criteria.   An advantage of the contracting aspects is that it is possible to avoid the multiverse and measure problems that arise in inflationary models.  Unlike ekpyrotic models, anamorphic models can be constructed using only a single field and can generate a nearly scale-invariant spectrum of tensor perturbations.  Anamorphic models also differ from pre-big bang and matter bounce models that do not explain the smoothness.  We present some examples of cosmological models that incorporate an anamorphic smoothing  phase.   
}

\begin{document}
\maketitle 
\flushbottom

\section{Introduction}
\label{sec:intro}

Observations show  the large-scale structure of the universe to be flat, smooth and scale-free~\cite{Ijjas:2013sua} and  the spectrum of primordial curvature perturbations to be nearly scale-invariant, adiabatic and Gaussian.  Inflationary cosmology attempts to explain these large-scale properties by invoking a period of rapid accelerated expansion, but this approach leads to an initial conditions problem and a multiverse problem that each make the theory unpredictive.  Ekpyrotic cosmology invokes a period of slow contraction that makes the universe smooth and flat while avoiding initial conditions and multiverse problems, but it requires an entropic mechanism using two or more scalar fields to generate the nearly scale-invariant spectrum of curvature perturbations.   

In this paper, we introduce ``anamorphic cosmology'' that combines elements and  advantages of both earlier approaches.  Based on some invariant criteria, the cosmological background during the smoothing phase behaves like a contracting universe that homogenizes, isotropizes and flattens  without introducing an initial conditions or multiverse problem; and, based on other invariant criteria, the cosmological background behaves like an expanding universe that can directly generate a scale-invariant spectrum of super-horizon adiabatic perturbations using a single scalar field.  Because the impression of the cosmological background depends on the perspective (i.e., the Weyl frame), we refer to models of this type as ``anamorphic.''

An anamorphic cosmology integrates a combination of concepts described in the forthcoming sections:  
\begin{itemize}[label=--]
\item  massive particles whose mass $m$ in any Weyl frame has a different time-dependence than the Planck mass $M_{\rm Pl}$, i.e., $m/M_{\rm Pl}$ is time varying (Sec.~\ref{sec:secessential});
\item   
two Weyl-invariant quantities,  $\Theta_m$ and $\Theta_{{\rm Pl}}$,  that characterize the contraction or expansion relative  to the characteristic matter scale (e.g., Compton wavelength) or to the gravitational (Planck) scale, respectively (Sec.~\ref{sec:secessential});
\item 
suppression of the anisotropy due to time-varying $m/M_{\rm Pl}$  (Sec.~\ref{sec:secbackground});
\item 
a scalar-tensor theory realization of the anamorphic phase (Sec.~\ref{sec:ScalarTensor});
\item 
a  kinetic coupling that causes  $\Theta_m$ and $\Theta_{\rm Pl}$  to have opposite signs (Sec.~\ref{sec:ScalarTensor});  
\item 
generation of nearly scale-invariant adiabatic, Gaussian {\it density} perturbations in an anamorphic contracting phase governed by a single scalar field (Sec.~\ref{sec:secperturbation});
\item 
generation of nearly scale-invariant {\it gravitational-wave} perturbations in an anamorphic contracting phase (Sec.~\ref{sec:secperturbation});
\item 
a novel, anamorphic mechanism for averting a multiverse (Sec.~\ref{sec:secnomult});  
\item 
two kinds of bounces (termed $\Theta_m$-bounce and $\Theta_{\rm Pl}$-bounce) in which $\Theta_m$ or $\Theta_{\rm Pl}$ reverse sign from negative to positive (Sec.~\ref{sec:secComplete});
\item  
transition  from contraction to expansion and vice versa without violation of the null-energy condition (NEC)   (Sec.~\ref{sec:secSimple});
\item  
or, in some cases, non-singular bounce from contraction to expansion with ghost-free, stable NEC violation (Sec.~\ref{sec:secCyclic});
\item 
incorporation of the elements above into a geodesically complete cyclic cosmology in which the anamorphic phase, matter-radiation creation, and dark-energy domination repeat at regular intervals (Sec.~\ref{sec:secCyclic}).   
\end{itemize}
Although some of these elements have been considered by other authors (most notably by Piao \cite{Piao:2011bz}, Wetterich \cite{Wetterich:2014eaa} and Li \cite{Li:2014qwa},  as we will discuss), the novelty here is how they work together with the other elements to form a complete cosmological scenario.

This paper is organized as follows. Secs.~\ref{sec:secessential}-\ref{sec:secperturbation} describe the anamorphic phase   during which the universe is smoothed and adiabatic curvature perturbations are generated.  In Secs.~\ref{sec:secessential}-\ref{sec:secadvantage},  we describe the essential components of anamorphic cosmology, the advantages compared to other cosmological scenarios, and the cosmological background. 
Sec.~\ref{sec:secbackground} details the conditions that the background solution must satisfy to smooth and flatten the universe and produce squeezed quantum states.
Then, in Sec.~\ref{sec:ScalarTensor}, we present a simple scalar-tensor field theory that realizes the anamorphic conditions.  In this paper, we focus on cases for which there exists a frame in which the anamorphic scalar field $\phi$ couples only to itself and the Ricci scalar; generalizations of  anamorphic cosmology will be discussed in forthcoming publications.  
The second-order action describing the perturbations and  the conditions for obtaining a nearly scale-invariant spectrum are given in Sec.~\ref{sec:secperturbation}.  The analysis shows how anamorphosis circumvents an earlier no-go theorem \cite{Creminelli:2004jg} stating that adiabatic modes always decay during contracting phases if mediated by a single scalar field.   We then explain how the multiverse is avoided. We conclude with examples that demonstrate how to incorporate an anamorphic smoothing phase into a complete cosmological scenario.   A simple model with a single anamorphic phase is analyzed in Sec.~\ref{sec:secSimple}.   In Sec.~\ref{sec:secCyclic}, we describe the construction of a model in which the smoothing phase, the bounce and the transition to a hot expanding phase repeat at regular intervals in a geodesically complete, cyclic cosmology.

\section{Essentials of anamorphic cosmology}
\label{sec:secessential}

Anamorphic cosmology relies on having two essential components during the smoothing phase: time-varying masses and a combination of Weyl-invariant signatures that incorporate aspects of contracting and expanding backgrounds. 

The first component is that particle masses and/or the Planck mass have different time dependence in any Weyl frame.
For simplicity, we will consider models in which matter-radiation consists of massive dust.   The frame-invariant action for a single particle is
\begin{equation} \label{Dustaction}
S_p = \int \frac{m}{M_{\rm Pl}}\, ds
\,,
\end{equation} 
where $d s$ is the line element and $m$ and $M_{\rm Pl}$ may each vary with time.   For a homogeneous and isotropic universe, we can assume a Friedmann-Robertson-Walker (FRW) metric $ds^2 = - dt^2 + a^2(t) dx_i dx^i$ where $t$ is the FRW time,  $a(t)$ is the scale factor and the Hubble parameter is $H \equiv d \, {\rm ln} \, a/dt$.

To distinguish cosmological models, it is convenient  to introduce the two dimensionless Weyl-invariant quantities:
\begin{eqnarray}
\label{thetas}
\Theta_{m}  & \equiv & \left( H+ \frac{\dot{m}}{m} \right) M_{\rm Pl}^{-1}
\,, \\
\Theta_{\rm Pl}  & \equiv & \left( H+ \frac{\dot{M}_{\rm Pl}}{M_{\rm Pl}} \right) M_{\rm Pl}^{-1}
\,.
\end{eqnarray}
Intuitively, $\Theta_{m}$ measures the physical expansion or contraction of the cosmological background as measured relative to a ruler (or any object made of matter); in the cases considered in this paper, $\Theta_{m}$ also measures the congruence of geodesics.   $\Theta_{\rm Pl}$ measures the evolution relative to the Planck mass, which is important for analyzing the generation of scalar and tensor metric perturbations. 

In standard cosmology with constant particle mass $m$ and Planck mass $M_{\rm Pl}$, the two invariants are equal,  $\Theta_{m} M_{\rm Pl}=\Theta_{\rm Pl}M_{\rm Pl}= H$, so no distinction is made.  In particular, both invariants are positive during an inflationary smoothing phase and negative  during an ekpyrotic smoothing phase.  One can introduce a Weyl transformation  of the metric in either case to re-express the action such that $M_{\rm Pl}$ or $m$ or both are time-dependent.  
In fact,  as Wetterich \cite{Wetterich:2014eaa} and Li  \cite{Li:2014qwa} have recently emphasized,  it is possible to transform expanding models to a frame where the Hubble parameter $H$ is negative.  However, the advantage of the variables $\Theta_{m}$ and $\Theta_{\rm Pl}$  is that they are frame invariant (resolving an ambiguity discussed in \cite{Domenech:2015qoa}).  For example, Wetterich's formulation of the  action can still be  unambiguously recognized as a version of expansion during all phases by computing these Weyl-invariant quantities and showing them both to be positive.  Analogous constructions are possible in ekpyrotic models, in which case both invariants are negative during the smoothing phase.

The second essential component of an anamorphic phase is that $\Theta_{m} $ is negative (as in ekpyrotic models) and $\Theta_{\rm Pl} $ is positive (as in inflationary models), which is possible in principle if $m$ and $M_{\rm Pl}$ are time-dependent, as can be seen from Eqs.~(\ref{thetas}).  In Sec.~\ref{sec:secbackground}, we will discuss how to construct models whose equation of state during the anamorphic phase leads to the pair of Weyl-invariant quantities having this behavior.  Because  $\Theta_{m} $ is negative, the cosmological background is physically contracting such that smoothing occurs without creating a multiverse or incurring an initial conditions problem, as in ekpyrotic models.  Because  $\Theta_{\rm Pl} $ is positive, the second-order action describing the generation and evolution  of curvature perturbations during the smoothing phase is similar to the case of inflation.  Consequently, a nearly scale-invariant spectrum of adiabatic density and gravitational wave perturbations can be generated in models with a single scalar field.  

From the invariants, therefore, it is clear that anamorphic models are unambiguously distinctive and, at the same time, share the best traits of inflationary and ekpyrotic cosmology while avoiding the worst.  The essential differences are summarized in Table I.   
\begin{table*}[t]
\begin{center}
\renewcommand\tabcolsep{10pt}
\renewcommand{\arraystretch}{1.3}
\begin{tabular}{|c|c|c|c|}\hline
& Anamorphosis & Inflation & Ekpyrosis\\
\hline\hline
$\Theta_m $&  $<0$ &$>0$&$<0$\\\hline
background& 
contracts& expands &  contracts \\
\hline \hline
$\Theta_{\rm Pl}$& $>0$ & $>0$ & $<0$\\\hline
curvature perturbations &grow  & grow  & decay\\
\hline
 \end{tabular}
\end{center}
\label{default}
\caption{Distinguishing the anamorphic, inflationary, and ekpyrotic scenarios: In conventional inflationary and ekpyrotic models, both the particle mass $m$ and the Planck mass $M_{\rm Pl}$ are constant, and the two Weyl-invariants $\Theta_m$ and $\Theta_{\rm Pl}$  are both equal to the Hubble parameter (in Einstein frame).  Consequently, both invariants are positive during an inflationary smoothing phase and both are negative during an  ekpyrotic smoothing phase. In anamorphic models,  $\Theta_m$, which characterizes the behavior of the cosmological background, is negative, as in ekpyrotic cosmology; and $\Theta_{\rm Pl}$, which characterizes the generation and evolution of scalar and tensor metric perturbations, is positive, as in inflationary models. 
}
\end{table*}

Table I suggests a fourth possibility with $\Theta_m$ positive and $\Theta_{\rm Pl}$ negative. In fact, such a set-up is achievable for different choices of  the equation of state and time-varying masses.  However, for the purposes of constructing useful cosmological models, this possibility is unpromising since it has the cosmological background behavior of inflation, thereby typically producing a multiverse, and the generation of curvature perturbations, governed by $\Theta_{\rm Pl}$, is made more complicated.

\section{Advantage over previous models}
\label{sec:secadvantage}

An anamorphic cosmology has several advantages over  the classic inflationary paradigm \cite{Guth:1980zm,Albrecht:1982wi,Linde:1981mu}.  First, unlike  inflation \cite{Penrose:1988mg,Gibbons:2006pa}, initial conditions do not have to be finely-tuned for the smoothing phase to begin.  Second, inflation is geodesically incomplete \cite{Borde:2001nh}, but cyclic anamorphic models can be constructed that are geodesically complete.   Third, eternal smoothing \cite{Steinhardt:1982kg,Vilenkin:1983xq} is avoided, so there is no multiverse.   Hence, the predictions in a given anamorphic model are unambiguous and testable.  

Among cosmological models that feature  periods of contraction, simple pre-big bang \cite{Veneziano:1991ek,Gasperini:1992em,Gasperini:1993hu,Gasperini:1994xg}  and matter bounce models \cite{Wands:1998yp,Finelli:2001sr} are unstable to anisotropy perturbations and, hence, do not explain smoothness and flatness.  Anamorphic and ekpyrotic models do, but they employ different smoothing mechanisms.  Ekpyrosis  \cite{Khoury:2001wf} uses  a period of ultra-slow contraction in which the pressure, $p$, exceeds the energy density, $\rho$  ($w\equiv p/\rho>1$).  The anamorphic model relies instead on time-varying masses to suppress the anisotropy and imposes different conditions on the equation of state, as described in the forthcoming sections. 

A disadvantage of ekpyrosis is that it is not possible to generate scale-invariant curvature perturbations with a single scalar field during the slow contraction phase \cite{Creminelli:2004jg}.  A second scalar field is needed   such that one linear combination of the two fields creates scale-invariant entropic perturbations that are converted to curvature perturbations prior to or during the bounce \cite{Notari:2002yc,Finelli:2002we,Lehners:2007ac,Buchbinder:2007ad}. By contrast, the anamorphic mechanism requires only a single scalar field  to do the smoothing, flattening and generation of curvature perturbations.  Entropic perturbations and a conversion from entropic to curvature perturbations are not needed. 

In the ekpyrotic picture,  there is no known way of generating a scale-invariant spectrum of tensor fluctuations.   The {\it primordial} tensor-to-scalar ratio is effectively $r = 0$, a firm prediction that can be proven wrong in future experiments.   (N.B. Secondary perturbations are automatically generated in this and all cosmological models by density fluctuations entering the horizon after matter domination; these contribute an effective $r\lesssim10^{-6}$ \cite{Baumann:2007zm}.)  The anamorphic model  generates a nearly scale-invariant spectrum of gravitational wave perturbations.  Future observations will determine if this is an advantage.

\section{Conditions for an anamorphic smoothing phase}
\label{sec:secbackground}

As described in Sec.~\ref{sec:secessential}, the cosmological background during an anamorphic smoothing phase is described by two Weyl-invariant quantities --  $\Theta_m < 0$ and $\Theta_{\rm Pl} > 0$ -- where the first measures contraction with respect to a physical ruler composed of matter and the second measures the expansion with respect to the Planck length, respectively.   These quantities can be expressed in terms of two additional frame-invariants:
\begin{eqnarray}  
\label{alphas0}
\alpha_m  &\equiv& a\, \frac{m}{M_{\rm Pl}^0},
\\
\label{alphas01}
\alpha_{\rm Pl}   &\equiv&   a\, \frac{ M_{\rm Pl}}{M_{\rm Pl}^0}, 
 \end{eqnarray}
where $a$ is the FRW scale factor, $m$ is the particle mass, $M_{\rm Pl}$ is the reduced Planck mass expressed in a given frame and $M_{\rm Pl}^0$ is the value of the reduced Planck mass in the frame where it is time independent, e.g., Einstein frame in scalar-tensor theories.
The invariant quantities 
$\alpha_m$ and $\alpha_{\rm Pl}$ satisfy the relations:
\begin{eqnarray}
\label{alphas}
\Theta_m &=& M_{\rm Pl}^{-1} \frac{\dot{\alpha}_m}{\alpha_m}  \quad {\rm and} \\
\label{alphas1}
\Theta_{\rm Pl} &=& M_{\rm Pl}^{-1} \frac{ \dot{\alpha}_{\rm Pl} }{ \alpha_{\rm Pl} },
\end{eqnarray}
so that  $\alpha_{m,\rm Pl}$  can be viewed as the effective Weyl-invariant scale factors associated with the Weyl-invariant Hubble-like parameters $\Theta_{m,\rm Pl}$.

During the anamorphic phase, it is most useful to express 
the first Friedmann equation  in a frame-invariant form using $\Theta_m$ and $ \alpha_m$:
\begin{equation}
\label{FriedmannEq1}
3\,  \Theta_m^2 \left(1 -  \frac{ d\ln \left( m/M_{\rm Pl} \right) }{ d\ln \alpha_m} \right)^2
= \frac{ \rho_{\rm A} }{ M_{\rm Pl}^4 }
+ \frac{\rho_m}{M_{\rm Pl}^4 }
- \left(\frac{m}{M_{\rm Pl}}\right)^2 \frac{\kappa}{\alpha_m^2} + \left(\frac{m}{ M_{\rm Pl} }\right)^6 \frac{\sigma^2}{\alpha_m^6}.
\end{equation}
The Friedmann equation describes the contributions of different forms of energy density  and curvature to the contraction or expansion rate.   The first contribution, $\rho_{\rm A}/M_{\rm Pl}^4 $, is due to the anamorphic energy density that dominates during the smoothing phase.  In the example given in the next section, $\rho_{\rm A}$ is ascribed to a non-minimally coupled scalar field $\phi$ with interaction potential $V_J(\phi)$.   The second contribution, $\rho_{\rm matter}/M_{\rm Pl}^4$, is due to the matter-radiation energy density in which the matter consists of particles with mass $m$.  The last two contributions are due to the spatial curvature, where $\kappa = (+1,0,-1)$,  and to the anisotropy, parameterized by $\sigma^2$.  

The anamorphic combination of $\Theta_m < 0 < \Theta_{\rm Pl}$ requires that $m$ and/or $M_{\rm Pl}$ be time-dependent such that the invariant mass ratio $m/ M_{\rm Pl}$ decreases with time.  
Note that the factors of $m/M_{\rm Pl}$ in the Friedmann equation suppress the spatial curvature and, even more so, the anisotropy. This is a key feature because suppressing the anisotropy in a contracting universe is essential for avoiding chaotic mixmaster behavior and maintaining homogeneity and isotropy.    
At some point after the anamorphic phase, to reach consistency with all current cosmological observations and tests of general relativity,
the universe must reheat to a high temperature and enter a hot expanding phase in which  both $m$ and $M_{\rm Pl}$  become constant, with  $M_{\rm Pl}=M_{\rm Pl}^0$,  the current value of the reduced Planck mass.
Throughout this paper, we will use reduced Planck units with $M_{\rm Pl}^0\equiv 1$, except where specified otherwise.

Of course, the Friedmann equation during the anamorphic phase can also be described using the invariants $\Theta_{\rm Pl}$ and $ \alpha_{\rm Pl}$, as we will do at the end of this section.  However, these variables are less useful because they obscure the role of the time-varying masses and the conditions needed after the anamorphic phase to join onto a hot expanding phase. 

Next, we consider  the generic conditions that $\Theta_m$ and $\alpha_m$ must satisfy to resolve the homogeneity and isotropy problems and to generate nearly scale-invariant perturbations.  

\textit{Smoothing condition.} 
To quantify the condition for smoothing in a model-independent way, we define the effective equation-of-state parameter, $\epsilon_{m}$, and a mass-variation index $q$ according to the following relations:
\begin{eqnarray}
\label{var}
\epsilon_m &\equiv& - \frac{1}{2} \frac{d\ln \Theta_m^2  }{d\ln\alpha_m},
\end{eqnarray}
and
\begin{eqnarray}
\label{p}
q &\equiv& \frac{d\ln \left( m/M_{\rm Pl} \right)}{d\ln \alpha_m}.
\end{eqnarray}
Substituting the second relation into Eqs.~(\ref{alphas0}-\ref{alphas01}), we find  
\begin{equation}
\label{transform}
\frac{d\alpha_{\rm Pl} }{ d \alpha_m} = 1-q.
\end{equation}  
Recalling that, $\alpha_m$ is decreasing and $\alpha_{\rm Pl}$ is increasing in an anamorphic phase,  we see that the mass-variation index must satisfy $q>1$. Note that this constraint implies the condition that the invariant mass ratio $m/M_{\rm Pl}$ is decreasing during the anamorphic phase.

In general, $\epsilon_m$ and $q$ can vary with time; however, for simple models, 
smoothing for an extended period of time is achieved by having them change very slowly during the anamorphic phase;  hence, treating them as roughly constant is a useful approximation deep in the anamorphic phase.  
In the case of nearly constant $q$, the expression for $\Theta_m$ takes a particularly simple form, namely
$\Theta_m^2 \propto 1/ \alpha_m^{ 2 \epsilon_m } $.

In analogy with ekpyrotic and inflationary models, the conditions for smoothing and flattening  an anamorphic universe
can be expressed as constraints on the equation-of-state $\epsilon_m$.  They require that the anamorphic energy density  $\rho_{\rm A}/M_{\rm Pl}^4 $  dominates all other contributions on the right hand side of the Friedmann equation for an extended amount of time.  ``Domination'' means $\Theta_m^2 (1-q)^2 \propto  \rho_{\rm A}/M_{\rm Pl}^4$ or, from Eq.~(\ref{var}), 
\begin{eqnarray}
\label{smooth}
 \frac{d\ln \left( \rho_{\rm A} /M_{\rm Pl}^4 \right)}{d\ln \alpha_m} = - 2\epsilon_m + \frac{d\ln (1-q)^2}{d\ln \alpha_m} .
\end{eqnarray}
Note that the second expression on the right hand side of Eq.~\eqref{smooth} is very small, in general, so $ \rho_{\rm A}/M_{\rm Pl}^4$ is roughly proportional to  $1/ \alpha_m^{ 2 \epsilon_m }$ .
In order for Eq.~\eqref{smooth} to hold as $\alpha_m$ shrinks during the anamorphic contracting phase, $2\epsilon_m $ must exceed the corresponding exponents for the spatial curvature and anisotropy terms in the Friedmann equation if they are expressed as powers of $1/ \alpha_m^{ 2 \epsilon_m }$.  
This condition yields a pair of constraints on $\epsilon_m$:
\begin{eqnarray}
\label{smcon}
 \epsilon_m   \gtrsim 1 - q \quad \& \quad \epsilon_m   \gtrsim 3(1-q). 
\end{eqnarray}
Here, for simplicity, we have neglected the weak time-dependence of $q$ (like the second term on the right hand side of Eq.~\eqref{smooth}). 
Since $q > 1$, both conditions are satisfied if the first inequality is satisfied, i.e., if $\epsilon_m \gtrsim 1-q$.   
(Note that an ekpyrotic contracting phase corresponds to  both $\alpha_m$ and $\alpha_{\rm Pl}$ decreasing and constant $m/M_{\rm Pl}$.  Hence,  $q=0$ and the smoothing constraint in Eq.~(\ref{smcon}) reduces to $\epsilon_m=\epsilon\equiv (3/2) (1+w)>3$, the known result.)

\textit{Squeezing of quantum curvature perturbations.}
To obtain a nearly scale-invariant spectrum of super-horizon curvature perturbations, the cosmological background must have the property that modes whose wavelengths are inside the horizon at the beginning of the smoothing phase can have wavelengths larger than the horizon size by the end of the smoothing phase.  The `horizon' is roughly the Hubble radius  $|H|^{-1}$ in standard ekpyrotic and inflationary cosmology, and the squeezing condition (modes exiting the horizon scale during smoothing) is that $|a H|$ is increasing.  

More generally, the `horizon' is a dynamical length scale that separates smaller wavelengths for which the curvature modes are oscillatory from the large wavelengths for which the curvature modes become frozen.  In anamorphic models, the evolution of scalar and tensor metric perturbations are described  entirely by the gravitational and scalar field parts of the effective action and do not depend on particle mass, as shown in Sec.~\ref{sec:secperturbation}.  Hence, the dynamical length scale is set by $\Theta_{\rm Pl}^{-1}$ and the corresponding squeezing condition  
is that $\alpha_{\rm Pl} \Theta_{\rm Pl\rm Pl} $ be increasing,  
\begin{equation}
\label{sq1}
\frac{d |\alpha_{\rm Pl}\Theta_{\rm Pl}|}{d\,t} M_{\rm Pl}^{-1}  > 0,
\end{equation}
which reduces to the standard condition in inflationary and ekpyrotic models. 
As shown in Appendix~\ref{appendix:Conditions}, if combined with the second Friedmann equation,
 the squeezing constraint reduces to the same condition as the smoothing constraint described above in Eq.~(\ref{smcon}).
Hence, squeezing imposes no additional constraint.

\textit{Sufficient anamorphosis.}
Finally, the anamorphic smoothing phase must last sufficiently long  to solve the horizon and flatness problems and generate a broad enough band of scalar and tensor metric perturbations that spans the observable horizon today.  Since the metric perturbations are most simply described by $\alpha_{\rm Pl}$ and $\Theta_{\rm Pl} $, as noted above,  it is most straightforward to label modes by the variable $N$, the number of $e$-folds of anamorphic phase remaining after the modes become frozen, as measured by the  increase in $\alpha_{\rm Pl}$:
\begin{equation}
\label{N}
N \equiv  \int_{\alpha_N}^{\alpha_0} d\,\ln \alpha_{\rm Pl} , 
\end{equation}
 where $\alpha_N = \alpha_{\rm Pl}(N)$ is the value of $\alpha_{\rm Pl}$ at the $N$th $e$-fold mark and $\alpha_0$ is the value at reheating.   $N$ runs from some large positive $N_{\rm max}$ to zero during the anamorphic phase.  The condition that the duration of the anamorphic phase is sufficient to smooth and flatten the universe is $N_{\rm max}>60$.   
 
\textit{Other relations.}
The expansion of the cosmological background relative to fixed rulers (set by the constant particle mass $m$) is most clearly described by the invariants $\alpha_m$ and $\Theta_m$, as we have done above.  However, it is always possible 
 to use  the invariants $\Theta_{\rm Pl}, \alpha_{\rm Pl}$, instead. Using these variables simplifies some expressions but has
the disadvantage that it obscures the effects of time-varying mass since $\alpha_{\rm Pl} = \alpha_m (M_{\rm Pl}/m)$ absorbs the mass dependence.  Also, because $\alpha_{\rm Pl}$ is increasing even though the universe is physically contracting, these variables are not useful for determining how to join properly the anamorphic contracting phase to a hot expanding phase.
For completeness, though, we repeat here many of the relations above that were expressed in terms of
 the invariants invariants $\alpha_m$ and $\Theta_m$.  
 
 The first Friedmann equation reduces to  
\begin{equation}
\label{FriedmannEq1Pl}
3\,  \Theta_{\rm Pl}^2 
= \frac{ \rho_{\rm A} }{M_{\rm Pl}^4 }
+ \frac{\rho_m}{M_{\rm Pl}^4 }
-  \frac{\kappa}{\alpha_{\rm Pl}^2} +  \frac{\sigma^2}{\alpha_{\rm Pl}^6}. 
\end{equation} 

In analogy with the definition of $\epsilon_m$ in Eq.~(\ref{var}), it is useful to introduce an effective equation-of-state parameter $\epsilon_{\rm Pl}$ such that 
\begin{eqnarray}
\label{var2}
\epsilon_{\rm Pl} &\equiv& - \frac{1}{2} \frac{d\ln \Theta_{\rm Pl}^2}{d\ln  \alpha_{\rm Pl}}.
\end{eqnarray}
Using the relation $(1-q) d \alpha_m = d\alpha_{\rm Pl} $ and the definition of the mass-variation index in Eq.~(\ref{p}), it is straightforward to show that
\begin{equation}
\label{epsmTOepsPl}
\epsilon_m  - \frac{1}{2}\frac{d\ln (1-q)^2}{d\ln \alpha_m} = (q-1) \epsilon_{\rm Pl},
\end{equation}
such that the smoothing and squeezing constraints reduce to
\begin{eqnarray}
\label{ePLc}
 \epsilon_{\rm Pl} < 1.
\end{eqnarray}

In a frame-invariant form, the second Friedmann equation is given as
\begin{equation}
\label{Hdot}
\frac{d \Theta_{\rm Pl}}{d\, t}M_{\rm Pl}^{-1} = -\frac{1}{2} \frac{\rho_A+p_A}{M_{\rm Pl}^4},
\end{equation}
where $p_A/M_{\rm Pl}^4$ is due to the anamorphic pressure. Combining with the first Friedmann equation in Eq.~\eqref{FriedmannEq1}, we find
\begin{equation}
\label{epsTOepsPL}
\epsilon_{\rm Pl} =  \frac{3}{2} \frac{\rho_A+p_A}{\rho_A};
\end{equation}
in particular, $\epsilon_{\rm Pl}$ is equivalent to the conventional equation of state parameter $\epsilon$.

 \section{Scalar-tensor formulation}
\label{sec:ScalarTensor}

The discussion in previous sections invoked Weyl-invariant variables that describe the contraction or expansion relative to time-varying particle mass and Planck mass.  This formulation can be applied to a wide range of models and metrics, as we will explore in future publications.  However, to make the ideas concrete, we will henceforth confine our discussion  to models in which the anamorphic phase and the transition to a conventional hot expanding phase are mediated by a single scalar field $\phi$ that is non-minimally coupled to the Ricci scalar $R$ and non-linearly coupled to its kinetic energy density.  

The action for our scalar-tensor examples can be expressed as
\begin{equation}
\label{Genaction}
S = \int d^4x \sqrt{-g}\left( \frac{1} {2}M_{\rm Pl}^2(\phi) R - \frac{1}{2}k(\phi) (\partial_{\mu}\phi)^2  - V_J(\phi) +{\cal L}_m(\phi)  \right),
\end{equation} 
where $g_{\mu\nu}$ is the metric; $R$ is the Ricci scalar; $k(\phi)$ is the non-linear kinetic coupling function;   $V_J(\phi)$ is the potential energy density; and ${\cal L}_m$ is a Lagrangian density that describes all other matter and radiation degrees of freedom. Henceforth, we will use the label $J$ to denote frame-dependent quantities in Jordan-frame representation; $E$ to denote Einstein frame; and no designation for quantities that are Weyl-invariant. 

In the simple examples considered in this paper,  there is a Weyl transformation such that ${\cal L}_m$ is independent of $\phi$ and there is another Weyl transformation that makes $M_{\rm Pl}$ independent  of $\phi$ and  ${\cal L}_m$ $\phi$-dependent.  
The corresponding equations of motion are given in Appendix \ref{appendix:background}.
$M_{\rm Pl}(\phi) \equiv M_{\rm Pl}^0 \sqrt{f(\phi)}$  is positive definite and approaches the current value $M_{\rm Pl}^0 \equiv 1$ following the anamorphic phase and  through the present epoch.   
These properties are  typical of general scalar-tensor theories.  The distinctive feature that enables  the anamorphic phase and the transition to and from is  the variation of the kinetic coupling, $k(\phi)$,  from positive (correct sign) to negative (wrong sign) or vice versa.  

The fact that $k(\phi)$ crosses through zero prevents its elimination by a field redefinition.  
In all cases, the action is constructed so that it is ghost-free.  In some scenarios, the action above suffices.  In others, we introduce additional Galileon-type terms, $\sim (\partial\phi)^2 \Box \phi, \sim (\partial\phi)^4$, that play a role before or after the anamorphic phase.  

We note that Piao \cite{Piao:2011bz} and Li \cite{Li:2014qwa} discussed scalar-tensor theories with negative kinetic couplings as a mechanism for generating nearly scale-invariant tensor fluctuations.  We will see a similar effect on tensor fluctuations in our anamorphic models. However, Piao's example is explicitly inflationary (in our classification, $\Theta_m= \Theta_{\rm Pl}>0$), and Li's  is ill-defined since he does not specify the properties of the matter sector, though,  he describes his model as dual to inflation.  Another difference with Li is that his model maintains the negative kinetic coupling at all stages, including after the reduced Planck mass becomes fixed, which can add some complications after the smoothing phase.)

As we show in  Appendix~\ref{appendix:Conditions}, the conditions for smoothing and flattening the universe and for generating scale-invariant scalar and tensor metric perturbations combined with the no-ghost condition can be re-expressed in terms of the functions $f(\phi), k(\phi)$ and the equation-of-state parameter $\epsilon_m$ or, using Eq.~\eqref{epsmTOepsPl}, $\epsilon_{\rm Pl}\equiv \epsilon$.  
Combining these  constraints, we have
\begin{equation}
\label{allconstraints}
0 < 3 + 2k(\phi)\frac{f(\phi)}{(M_{\rm Pl}^0 f,_{\phi})^2}  < \epsilon  < 1.
\end{equation}
Here, the left-hand inequality is the no-ghost condition $(K(\phi) \equiv (3/2)(M_{\rm Pl}^0 f,_{\phi}/f)^2 + k/f >0)$; the middle inequality is the contraction condition ($\Theta_m<0$); and the third inequality is the smoothing and squeezing constraint ($\epsilon <1$). 

Clearly,  given that $f(\phi)>0$, Eq.~\eqref{allconstraints} can only be satisfied if the kinetic function 
\begin{equation}
\label{allconstraints2}
k(\phi) < 0
\end{equation}
during the smoothing phase.  As we have emphasized, $k(\phi)<0$ does not cause any physical instability provided the no-ghost condition, $K(\phi)>0$, is satisfied, which is already incorporated in Eq.~(\ref{allconstraints}).  

Finally, with $d\ln \alpha_{\rm Pl} = \Theta_{\rm Pl}\, (M_{\rm Pl}/\dot{\phi}) \,d\phi$ and the expression for $\epsilon$ in Eq.~(\ref{epsilon}), we can rewrite the expression for $N$ in Eq.~\eqref{N} as
\begin{equation}
\label{N2}
N = \int_{\phi_N}^{\phi_0} 
 \sqrt{\frac{K(\phi)}{2\epsilon}} \frac{d\phi}{M_{\rm Pl}^0}
 \gtrsim 60, 
\end{equation}
such that the condition of sufficient anamorphosis translates into the constraint that Eq.~(\ref{allconstraints}) holds for $\phi_{60}\leq \phi< \phi_0$.
These  two conditions on $f(\phi), k(\phi)$ and $\epsilon$  are sufficient to have an anamorphic smoothing phase that produces a homogeneous and isotropic cosmological background.  
As in the case of inflationary and ekpyrotic models, stronger constraints must be imposed in order for the spectrum of perturbations to be nearly scale-invariant, as discussed in the next section.

\section{Perturbations}
\label{sec:secperturbation}

In this section, we discuss the  generation of anamorphic scalar and tensor fluctuations and show that adiabatic curvature modes with a (nearly) scale-invariant spectrum are generated.  Notably, the generation mechanism requires only a single field,  is dynamically stable, results in negligible non-Gaussianity, and produces tensor modes in addition to scalar modes.   These are all features that do not occur in ekpyrotic models and have been  considered unattainable in bouncing cosmologies \cite{Creminelli:2004jg,Tolley:2007nq,Koyama:2007if,Buchbinder:2007at,Lehners:2007wc,Lehners:2008my}.  The key reason why they can occur in the anamorphic scenario is that, even though the parameter $\Theta_m$ is negative so that the physical background is contracting, $\Theta_{\rm Pl}$ is positive.   A related effect is that the spectral tilts of both the scalar and tensor perturbation spectra are typically red, whereas one might expect them to be blue in a contracting universe.  We will also explain why the anamorphic scenario has no multiverse problem, which is more subtle than the case in ekpyrotic models.

\subsection{Adiabatic and tensor modes from the anamorphic phase}
\label{sec:sec4.1}

The linearized perturbations of our model can be evaluated using the second order actions for both scalar curvature and tensor perturbations. 

Curvature perturbations, $\zeta$, are local perturbations in the scale factor that are described by the perturbed metric
\begin{equation}
ds^2 = -dt^2 + a(t)e^{2\zeta(t, x^i)}dx^idx_i .
\end{equation}
Following the standard calculation \cite{Bardeen:1983qw,Mukhanov:1990me}, we obtain the (second-order) action for scalar curvature perturbations $\zeta,$
\begin{equation}
S_{2,S} = \frac{1}{2} \int d \eta d^3 x \, \alpha_{\rm Pl}^2 \epsilon_{\rm Pl} \left[ \left( \zeta_{,\eta} \right)^2 - \left( \partial_k \zeta \right)^2 \right], \label{eq:S2scalar}
\end{equation}
where both $\alpha_{\rm Pl}$ and $\epsilon_{\rm Pl}$ are defined as in Eqs.~(\ref{thetas}) and (\ref{var}) and we have switched to conformal time $\eta$, given by $d\eta = d\,t/a$. Note that $\eta$ is frame invariant.

Defining a transverse and traceless tensor perturbation $h_{ij}$ by
\begin{equation}
\delta g_{ij} \equiv a^2 h_{ij},
\end{equation}
we obtain the analogous quadratic action for tensor perturbations,
\begin{equation}
S_{2,T} = \frac{1}{8} \int d\eta d^3 x \, \alpha_{\rm Pl}^2 \left[ \left( h_{ij,\eta} \right)^2 - \left( \partial_k h_{ij} \right)^2 \right]. 
\label{eq:S2tensor}
\end{equation}
From the definition of the curvature and tensor perturbations and Eq.~(\ref{EFmetric}), it follows that both $\zeta$ and $h_{ij}$ are frame invariant as are the prefactors $\alpha_{\rm Pl}^2, \alpha_{\rm Pl}^2 \epsilon_{\rm Pl}$.
Hence, the perturbed quadratic actions are invariant under a frame change.  As anticipated in earlier sections, although the cosmologically contracting background is most intuitively described in terms of $\alpha_m$, the perturbations are most intuitively described by $\alpha_{\rm Pl}$.

In order to obtain nearly scale-invariant spectra of curvature and tensor perturbations, we must choose the couplings, $f(\phi), k(\phi)$ and the potential, $V_J(\phi)$, such that the corresponding pre-factors $ \alpha_{\rm Pl}^2 \epsilon_{\rm Pl}$ and $\alpha_{\rm Pl}^2$ scale approximately as $\eta^{-2}$, where any small correction to the exponent leads to a tilt.  

In an ekpyrotic contracting phase, there is no such option.   The pre-factors in the corresponding quadratic actions behave differently: for scalar modes, the pre-factor is proportional to $\alpha_{\rm Pl}^2\equiv \alpha_m^2$ (assuming a constant or slowly varying equation of state $\epsilon_{\rm Pl}\equiv \epsilon_m$) where $\alpha_m$ is very slowly decreasing as the universe contracts. This behavior is clearly different from $\eta^{-2}$, which is rapidly growing.  This explains why neither tensor modes nor adiabatic scalar modes get amplified during an ekpyrotic phase for models with a single scalar field. With two-fields, there is a mechanism to generate nearly scale-invariant adiabatic perturbations by first generating isocurvature perturbations and the converting them, but there is no analogous mechanism for tensor modes  so they are predicted to be absent  \cite{Lehners:2007ac,Buchbinder:2007ad}.  

In the anamorphic contracting phase, the situation is different because the pre-factors depend on $\alpha_{\rm Pl} \neq \alpha_m$, which is rapidly growing even though the effective scale factor $\alpha_m$ associated with the cosmological background (as measured by $\Theta_m$) is decreasing.  In the limit that $\epsilon_{\rm Pl}$ is nearly constant and much smaller than unity,  $\alpha_{\rm Pl}^2  \epsilon_{\rm Pl}$ and $\alpha_{\rm Pl}^2$ are nearly proportional to $\eta^{-2}$ and the spectrum is nearly scale invariant. In Sec.~\ref{sec:secExamples}, we shall show that it is possible to construct such examples.

We note that the background constraints (smoothing, flattening and squeezing) only require $\epsilon_m>1-q$ or  $\epsilon_{\rm Pl} \equiv \epsilon <1$, but nearly scale-invariant perturbations require stronger conditions, $\epsilon_m \approx 0$ or $ \epsilon_{\rm Pl} \ll 1$ and nearly constant over tens of $e$-folds of smoothing.  The same comment applies to inflation and ekpyrotic models.  

It is a common myth that scale-invariance is an automatic outcome of smoothing and flattening. For example, it is sometimes described as a prediction of inflation.  In none of these cosmologies is this actually the case.  Near scale-invariance, in accordance with observations, imposes significantly tighter constraints on the equation-of-state and its time variation.  In the case of inflation, for example, the condition is imposed by requiring that the inflationary phase be nearly de Sitter, which is much stronger than needed to smooth and flatten.

\subsection{No multiverse}
\label{sec:secnomult}

The multiverse arises in inflation because of the same quantum physics  that produces the scale-invariant spectrum described in the previous subsection.  Rare quantum fluctuations keep some regions of space in the smoothing phase when a naive estimate would suggest that inflation should have ended.  Within instants, the rare regions comprise most of the volume of the universe.   The situation repeats eternally,  
so that only a fractal volume less than three dimensions ever ends inflation.
This volume of measure zero containing regions that are no longer inflating is comprised of patches that complete inflation at different times after different random quantum fluctuations have affected the trajectory of the inflaton field that controls the rate of inflation.  The different fluctuations lead to different cosmological outcomes. In addition to patches that are flat, smooth, and have nearly scale-invariant perturbation spectra, there are infinitely many patches that are curved, inhomogeneous, anisotropic  and with non-scale-invariant fluctuations.  By spanning all possible cosmological outcomes, the multiverse effect makes the inflationary scenario unpredictive.  There is no generally accepted solution to this problem within inflationary cosmology at present \cite{Guth:2013sya}.

The multiverse problem does not occur in ekpyrotic cosmology because regions continuing to smooth due to rare fluctuations are contracting rather than expanding.  These rare contracting regions shrink away compared to typical regions that complete the contraction phase and transition to an expanding phase.  

A similar conclusion holds for anamorphic cosmology, but a more subtle analysis is required.   As we have emphasized, during the anamorphic smoothing phase the cosmological background is  contracting ($\Theta_m<0$) as measured by physical rulers comprised of particles with fixed mass $m$.  For rare fluctuations of the scalar field that create patches with negative $\Theta_m<0$ (within which  contraction continues for longer periods compared to typical patches), the situation is the same as in the ekpyrotic case. These rare regions shrink away compared to typical ones.  However, as we will illustrate with our simple example in Sec.~\ref{sec:secComplete}, there can also be rare fluctuations that stop or reverse the evolution of the scalar field and that can flip the sign of $\Theta_m$ to positive.  Now the rare patch is expanding rather than contracting, which cannot occur in the ekpyrotic picture.   We will see that these rare patches are not problematic, though.  There is a strong attractor solution that rapidly re-reverses the evolution of the scalar field and the sign of $\Theta_m$ before any significant expansion can occur.  After the reversal to contraction, the analysis reduces to cases already considered:  the rare patches never grow to dominate and there is no multiverse.

\section{Examples}
\label{sec:secExamples}

In this section, we illustrate some basic properties of anamorphic models using simplistic examples that are only intended to describe the anamorphic contracting phase.  In the next section, we describe how to embed such examples in models that describe a complete bouncing cosmology that includes a hot expanding phase following the anamorphic phase.

Let us consider the anamorphic action,
\begin{equation}
S = \int d^4x \sqrt{-g_J}\left( \frac{1}{2}\xi e^{2A\phi}R_J + \frac{1}{2}e^{2A\phi}(\partial_{\mu}\phi)^2 - V_J(\phi) \right),
\end{equation}
where we set the gravitational coupling $f(\phi) = \xi e^{2 A\phi}$ and the kinetic coupling $k(\phi) = -e^{2 A\phi}$ with both parameters $\xi$ and $A$ being positive real numbers and we set $M_{\rm Pl}^0=1$. 
For this family of models, the condition for anamorphic smoothing from Eq.~(\ref{allconstraints}) reduces to the simple form
\begin{equation}
\label{backgrExample}
0< \frac{6A^2\xi-1}{2\xi A^2} < \epsilon < 1.
\end{equation}
Recall that requiring strict positivity eliminates ghosts despite the fact that $k<0$. 

\subsection{Approximate solutions for $\epsilon(N)$}
\label{sec:secepsN}
If the smoothing condition in Eq.~\eqref{backgrExample} is satisfied, we can neglect the contributions from curvature, anisotropy, matter, radiation, etc. and, using the dimensionless ``time variable'' $N$ defined in Eq.~(\ref{N}), the equations of motion as given in Appendix~(\ref{JFeqsofmotion}-\ref{fieldeq0}) reduce to 
\begin{eqnarray}
\label{epsN}
\frac{\epsilon,_N}{\epsilon} &=& - 2\frac{\epsilon,_{\phi}}{\sqrt{2\epsilon K(\phi)}} 
= 2\frac{3 - \epsilon }{\sqrt{2\epsilon K(\phi)}} \left( \frac{V_J,_{\phi}}{V_J} - 2\frac{f,_{\phi}}{f} + \sqrt{2\epsilon K(\phi)} \right)
 \nonumber\\
&=&
  2(3 - \epsilon ) \left( \frac{1}{\sqrt{2\epsilon}} \sqrt{\frac{\xi}{6\xi A^2-1}} \left( \frac{V_J,_{\phi}}{V_J} - 4A\right) + 1 \right)
  . 
\end{eqnarray}
The solution for $\epsilon$ can be approximated analytically if we assume a slowly varying equation of state, i.e., 
\begin{equation}\label{slowN_def}
\frac{\epsilon,_N}{\epsilon} \Delta N < 1.
\end{equation}
To satisfy this inequality and the constraint that $\epsilon<1$ in Eq.~(\ref{backgrExample}), the factor in parentheses in Eq.~(\ref{epsN}) must be very small, or equivalently,
\begin{equation}\label{SlowRoll1}
 \epsilon \simeq \frac{1}{2 K(\phi)}  \left( \frac{d}{d\phi} \ln \left(V_J(\phi)/f^2(\phi)\right) \right)^2 
= \frac{1}{2} \frac{\xi}{6\xi A^2-1} \left( \frac{V_J,_{\phi}}{V_J} - 4A  \right)^2
.
\end{equation}

As noted at the end of Sec.~\ref{sec:sec4.1},  a slowly varying equation of state in the anamorphic (or inflationary or ekpyrotic) scenario is not necessary to solve the horizon and flatness problems. For example, it suffices that $\epsilon,_N \Delta N < 1 - \epsilon$ in the anamorphic and inflationary models.  Rather, it is the empirical constraint that the spectral tilt is observed to differ by only a few per cent from exact scale-invariance that forces the condition that $\epsilon$ be so slowly varying and $\ll 1$.  A similar statement applies to inflationary and ekpyrotic models.

For simplicity of presentation, we consider a sequence of potentials that can be studied analytically,
\begin{equation}
\label{VJexample}
V_J(\phi) = V_0 e^{B\phi- C \phi^p}, 
\end{equation} 
for constant $V_0$, $B$ and $C$ and integer $p$, although the results can be easily extended to potentials of the form 
 $V_J \propto e^{g(\phi)}$. 

\subsection{$V_J(\phi) = V_0 e^{B \phi}$}
\label{subsec:simplestPotential}

First, let us consider the case with $C=0$ and
\begin{equation}
\label{VJsimplest}
V_J(\phi) = V_0 e^{B \phi}.
\end{equation}
The pre-factor $V_0$ is a positive real number; it determines the amplitude of scalar perturbations, 
\begin{equation}
\label{amplitude}
\delta \rho/\rho(\phi_N) \sim \sqrt{K}\,\frac{\sqrt{V_J}}{f}\left( \frac{V_J,_{\phi}}{V_J} - 2\frac{f,_{\phi}}{f} \right)^{-1} 
\sim \sqrt{V_0} \sqrt{\frac{6\xi A^2-1}{\xi}} \frac{e^{(1/2) (B-4A)\phi_N}}{\xi |B-4A|}  . 
\end{equation}
Using Eq.~(\ref{epsN}), it is easy to see that the potential in Eq.~(\ref{VJsimplest}) corresponds to a constant equation of state,
\begin{equation}
\label{constEps}
\epsilon \equiv \frac{1}{2} \frac{\xi}{6A^2\xi-1} (B-4A)^2.
\end{equation}
The condition for anamorphic contraction in Eq.~(\ref{backgrExample}) reduces to
\begin{equation}
\label{backgrExample1}
0<  \frac{1}{2} (4A - B)^2 < \frac{6A^2\xi-1}{\xi} < A |4A-B|,
\end{equation}
where the first inequality is the no-ghost condition ($K(\phi)>0$), the second inequality is the smoothing condition ($\epsilon < 1$), and the third inequality is the condition for contraction ($\Theta_m <0$). Note that, in this example, $|4A-B|$ must be strictly greater than zero and non-negligible.

For the purposes of illustration, we introduce a modification that suddenly ends the smoothing phase at $\phi=\phi_0$.  (We will present a practical example of this in the next subsection.) 
Then, the number of $e$-folds of anamorphic contraction remaining is given by Eq.~(\ref{N2}),
 \begin{equation}
\label{Nsimple1}
N =  \frac{6A^2\xi-1}{\xi} \frac{1}{|4A-B|} \Delta\phi,
\end{equation}
where $\Delta\phi\equiv \phi_0 -\phi_N>0$.  
Using Ref.~\cite{Wang:1997cw}, the expression for the scalar tilt immediately follows,
\begin{equation}
\label{constEpsTilt}
n_S-1 = -2\epsilon + \frac{d \ln\epsilon}{d\,N} = - 2\epsilon = \frac{\xi}{6A^2\xi-1} (B-4A)^2 = -\frac{|4A-B| \Delta\phi}{N}.
\end{equation}
Observations of the CMB dictate that $n_S-1 \simeq - 1/30$ at $N \approx 60$, and, thus, for this example, $\Delta\phi \simeq 2/|4 A -B|$ and $(6A^2\xi-1)/\xi \simeq (4A-B)^2(N/2)$. 
Note that, with this choice of parameters, both the smoothing and the no-ghost constraints in Eq.~(\ref{backgrExample}) are fulfilled. Contraction occurs if, in addition, we require that $|4A-B| (N/2)<A$.  

The result demonstrates that, in principle, anamorphic contracting models can generate a nearly scale-invariant spectrum of tensor fluctuations.   

\subsection{$V_J(\phi) = V_0 e^{B \phi - C\phi^p}$}
\label{subsec:simplePotential}

A simple modification of the previous example is $V_J = V_0 e^{g(\phi)}$ where $g(\phi)$  is positive and increases linearly with $\phi$ for  small $\phi$ and is negative and decreasing for large $\phi$.  
As a specific example, we consider 
\begin{equation}
\label{VJsimple}
V_J(\phi) = V_0 e^{B\phi-C\phi^p}, \quad p \gg 2.
\end{equation} 
so that the anamorphic phase comes to an end when $\phi$ grows sufficiently large.
The equation of state $\epsilon$ begins very small and increases during the anamorphic phase,  reaching $\epsilon =1$ at $\phi=\phi_0\lesssim 1$  when the 
anamorphic phase ends.  

As in the previous example, the pre-factor $V_0$ determines the amplitude of scalar perturbations, 
\begin{equation}
\label{amplitude2}
\delta \rho/\rho(\phi_N) \sim  \sqrt{V_0 \frac{6\xi A^2-1}{\xi}} \frac{e^{(1/2) (B - 4A)\phi_N - (1/2) C\phi_N^p}}{\xi |B-4A - Cp\phi_N^{p-1}|}.
\end{equation}
The $\phi$-dependent equation of state is given by
\begin{equation}
\label{Eps2}
\epsilon \simeq \frac{1}{2} \frac{\xi}{6A^2\xi-1} \left( 4A - B + C\,p \phi^{p -1}\right)^2.
\end{equation}
Note that the equation of state is monotonically growing with increasing $\phi>0$ if the no-ghost condition, $(6A^2\xi-1)/\xi >0$, is satisfied.  This ensures that the density-fluctuation spectrum has a red tilt.   The anamorphic phase ends at  $\epsilon(\phi_0)=1$, so  
\begin{equation}
\label{phi0}
\phi_0 = \left( \left( \sqrt{2\,\frac{6A^2\xi-1}{\xi}} - 4A + B\right) \frac{1}{C\,p}\right)^{1/(p-1)}.
\end{equation}
Note that  $\epsilon<1$ and the smoothing condition is fulfilled for $0 < \phi<\phi_0$.

During the anamorphic phase, there are two regimes of interest: (I)  $|4A - B| \gtrsim C\,p\, \phi_0^{p-1}\gg C\,p\, \phi^{p-1}$; and (II)    $C\,p\, \phi^{p-1} \gg |4A-B|  \simeq  0 $:

\noindent
{\bf Case I}: If $|4A-B| \gtrsim C\,p\, \phi_0^{p-1}\gg C\,p\, \phi^{p-1}$, the term with coefficient $C$ in the potential in Eq.~(\ref{VJsimple}) is negligible and $\epsilon$ is nearly constant except just before the anamorphic phase ends when $\phi$ is very close to $\phi_0$, where  $ \phi_0$ is given in Eq.~(\ref{phi0}).  For $\phi_0\sim1$ and $N\sim 60$, the predictions are 
\begin{eqnarray}
n_S-1&  \simeq  &- |4A - B|/N, \\
 r& \sim& -8(n_S-1),
 \end{eqnarray}
  where  the contraction condition can be satisfied by choosing $A > N$. Note that, if the parameters are fit to the current observational constraint on the tilt, $n_S-1 \sim - 1/30$ at $N \approx 60$,  then Case I cannot satisfy the current observational bound on $r<10\%$ \cite{Ade:2015lrj}.
  
\hspace{0.01in}

\noindent
{\bf Case II}: If $C\,p\, \phi^{p-1} \gg |4A - B|  \simeq  0 $, the condition for contraction ($\Theta_m<0$) is given by
\begin{equation}
\label{backgrExample2}
0< \frac{6A^2\xi-1}{\xi} < A\, C\,p\, \phi_N^{p-1},
\end{equation}
where, in defining the upper bound, we have used the fact that $0<\phi_N<\phi_0$, i.e., 
if the physical background contracts for $\phi_N>0$, it contracts for all $\phi>\phi_N$.

For the number of $e$-folds of anamorphic contraction, we find 
\begin{equation}
\label{Nsimple2}
N =  \frac{6A^2\xi-1}{\xi} \frac{1}{C\,p} \int^{\phi_0}_{\phi_N} \frac{d\phi}{\phi^{p-1}} =  \frac{6A^2\xi - 1}{\xi} \frac{1}{C\,p} 
 \frac{1}{(p-2)} \left( \frac{1}{\phi_N^{p-2}} - \frac{1}{\phi_0^{p-2}} \right) ; 
\end{equation}
and the expression for the scalar spectral tilt is
\begin{equation}
\label{Tilt2}
(n_S-1)(\phi_N) = -2\, \frac{\xi}{6A^2\xi-1} C\,p(p-1) \phi_N^{p-2} \left( \frac{C\,p}{p-1} \phi_N^p +1 \right).
\end{equation}

Since $(\phi_0/\phi_N)^{p-2}\gg 1$, we have  $(n_S-1)(\phi_N) \simeq - (2/N) (p-1)/(p-2) (1+ C\,p\,\phi_N^p/(p-1))$.   In order to obtain the observed value of the tilt, $n_S - 1$,    $p$ must be large enough that $(p-1)/(p-2)  \sim 1$. 
To satisfy the remaining conditions for successful anamorphosis as stated in Eqs.~(\ref{phi0}-\ref{Nsimple2}), it suffices to choose the parameters $A, C, \phi_0$, and $\xi$ as follows,
\begin{equation}
\label{parameters}
A > \frac{(p-2)N}{\phi_N}; \; C\ll \frac{1}{\phi_N};\; \phi_0 = \left( 2\,\frac{p-2}{C\,p} N  \right)^{\frac{1}{2(p-1)}}\sqrt{\phi_N};\quad \frac{6A^2\xi-1}{\xi} = \frac{1}{2}C^2p^2\phi_0^{2(p-1)}
. 
\end{equation}
Using these parameter constraints, the equation of state and its derivative are given by
\begin{eqnarray}
\epsilon(\phi_N) & \simeq & \frac{\xi}{2 (6\xi A^2 -1)} C^2 p^2 \phi_N^{2 (p-1)} \simeq \left (\frac{\phi_N}{\phi_0} \right)^{2(p-1)} \ll \frac{2}{N} \phi_N^{p-2}, \\
\frac{d \, {\rm ln} \, \epsilon}{d N} (\phi_N) & \simeq & -\frac{2 \epsilon,_{\phi}}{\sqrt{2 \epsilon K(\phi)}} \simeq - \frac{2}{N}  .
\end{eqnarray}
Note that, since $\phi_N<<1$ and $p>2$,   $\epsilon$ is small compared to its derivative, $|d\ln\epsilon/dN|$. As a result, unlike Case I, it is possible to set the tilt $n_S-1 \sim d {\rm ln} \, \epsilon/d N$ to the observed value and have  the tensor-to-scalar ratio, $r\sim 16\epsilon$  be $\mathcal{O}(1/N^2)$ or smaller, which satisfies the current observational constraint on $r$.

\section{Completing the cosmological scenario}
\label{sec:secComplete}

As exemplified in the previous section, the anamorphic phase provides a smoothing mechanism that renders the universe homogeneous, isotropic and spatially flat, generating nearly scale-invariant, super-horizon curvature perturbations. 
For a complete cosmological scenario, though, we must also explain how the anamorphic phase ends  and smoothly connects to standard expanding big-bang evolution.  This includes specifying how $f(\phi)$ and $k(\phi)$ must behave during the transition.  

During the anamorphic phase, $\Theta_m<0$ and $\Theta_{\rm Pl}>0$.  In the expanding big-bang phase, $\Theta_m = \Theta_{\rm Pl}>0$.  To connect the two phases,  it is necessary that $\Theta_m$ switch sign.  We will use the term {\it $\Theta_m$-bounce} to refer to the moment when $\Theta_m$ switches sign from negative to positive; that is, a transition from $\alpha_m$ decreasing to $\alpha_m$ increasing.  (A switch from positive to negative $\Theta_m$, which corresponds to $\alpha_m$ switching from increasing to decreasing, will be called a {\it $\Theta_m$-reversal}.  The analogues for $\Theta_{\rm Pl}$ and $\alpha_{\rm Pl}$ will be called {\it $\Theta_{\rm Pl}$-bounce} and {\it $\Theta_{\rm Pl}$-reversal}.) 
      The first model we will describe in Sec.~\ref{sec:secSimple} is a {\it simple (one-time) scenario} that: (i) begins with an anamorphic smoothing phase;  (ii) undergoes a $\Theta_m$-bounce and reheats; and (iii) transitions to standard hot big-bang cosmology with decelerating expansion.  As with all current inflationary models, this scenario is incomplete because it does not include an explanation of what precedes the smoothing phase. The anamorphic phase only occurs once and the universe expands forever in the future.  $\Theta_{\rm Pl}$ is positive and monotonically decreasing throughout.  This simple model is sufficient to explain smoothness, flatness, and the generation of a nearly scale-invariant spectrum of density perturbations without inflation and without producing a multiverse.

The second example we will describe in Sec.~\ref{sec:secCyclic} is a {\it cyclic anamorphic scenario} with repeated epochs controlled by a single scalar field that each include:  (i) a dark energy dominated expanding phase (like the present universe) in which the field value is near zero; (ii) decay to a contracting phase during which both $\Theta_m$ and $\Theta_{\rm Pl}$ undergo a reversal and become negative; (iii) a $\Theta_{\rm Pl}$-bounce in which $\Theta_{\rm Pl}$ switches from negative to positive via a novel well-behaved NEC-violating phase; (iv) an anamorphic smoothing phase; (v) a $\Theta_m$-bounce with reheating followed by a rapid return of the scalar field to its original value (near zero); (vi) normal hot big-bang expansion until the dark energy overtakes matter and radiation, which brings the cycle back to step (i).    In principle, the cycles can continue forever following this sequence.  The cyclic scenario has all the advantages of the simple model and also has the property that it can be made geodesically complete and thereby  avoid initial condition problems.

\subsection{Simple (one-time) anamorphosis}
\label{sec:secSimple}
\begin{figure}[!tb]
\centering
\includegraphics[width=8.75cm]{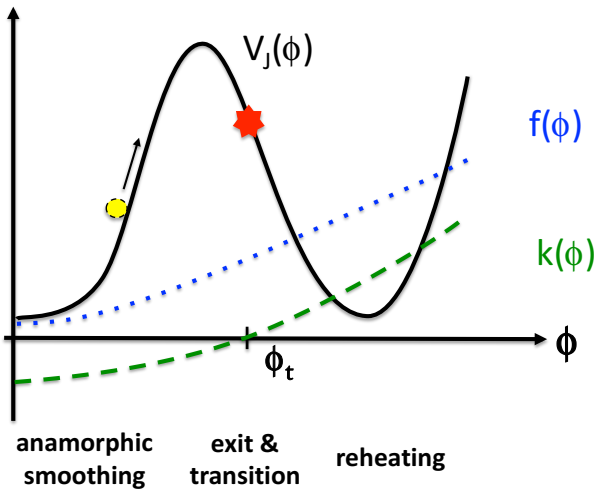}
\caption{Sketch of the (Jordan-frame) potential, $V_J$ (continuous black line), the gravitational coupling, $f$ (dotted blue line)  and the kinetic coupling, $k$ (dashed green line) versus $\phi$ in the simple anamorphic scenario.  During the smoothing phase,  the anamorphic field $\phi$ moves uphill along  the potential energy curve due to the wrong-sign kinetic term while $f$ grows monotonically.
The anamorphic phase ends when, on the way uphill, the field's kinetic energy becomes comparable with its potential energy and $\epsilon \rightarrow 1$.   The remaining  kinetic energy of the field is sufficient to carry $\phi$ further uphill  and over the top of the potential.  On the way downhill, a transition to expansion occurs ($\Theta_m$-bounce) when the kinetic coupling changes sign at  $\phi_t \sim M_{\rm Pl}^0$. After the $\Theta_m$-bounce, $\phi$ reaches the minimum of the potential well and starts to oscillate rapidly, decaying and reheating the universe. Near the minimum, $\phi$ obtains a large mass and $f$ and $k$ become fixed so that the Einstein and Jordan frames become indistinguishable.  The vacuum energy density at the minimum has a small positive value corresponding to the current dark energy density.
} 
\label{fig:1}
\end{figure}
In this subsection, we describe a simple (non-cyclic) anamorphic scenario in which the universe begins in the anamorphic smoothing phase and subsequently reheats and transitions to a standard big-bang phase.  Beginning in the anamorphic phase means that the condition in Eq.~(\ref{slowN_def}) is satisfied initially. %
\begin{figure}[tb]
\centering
\includegraphics[width=8.5cm]{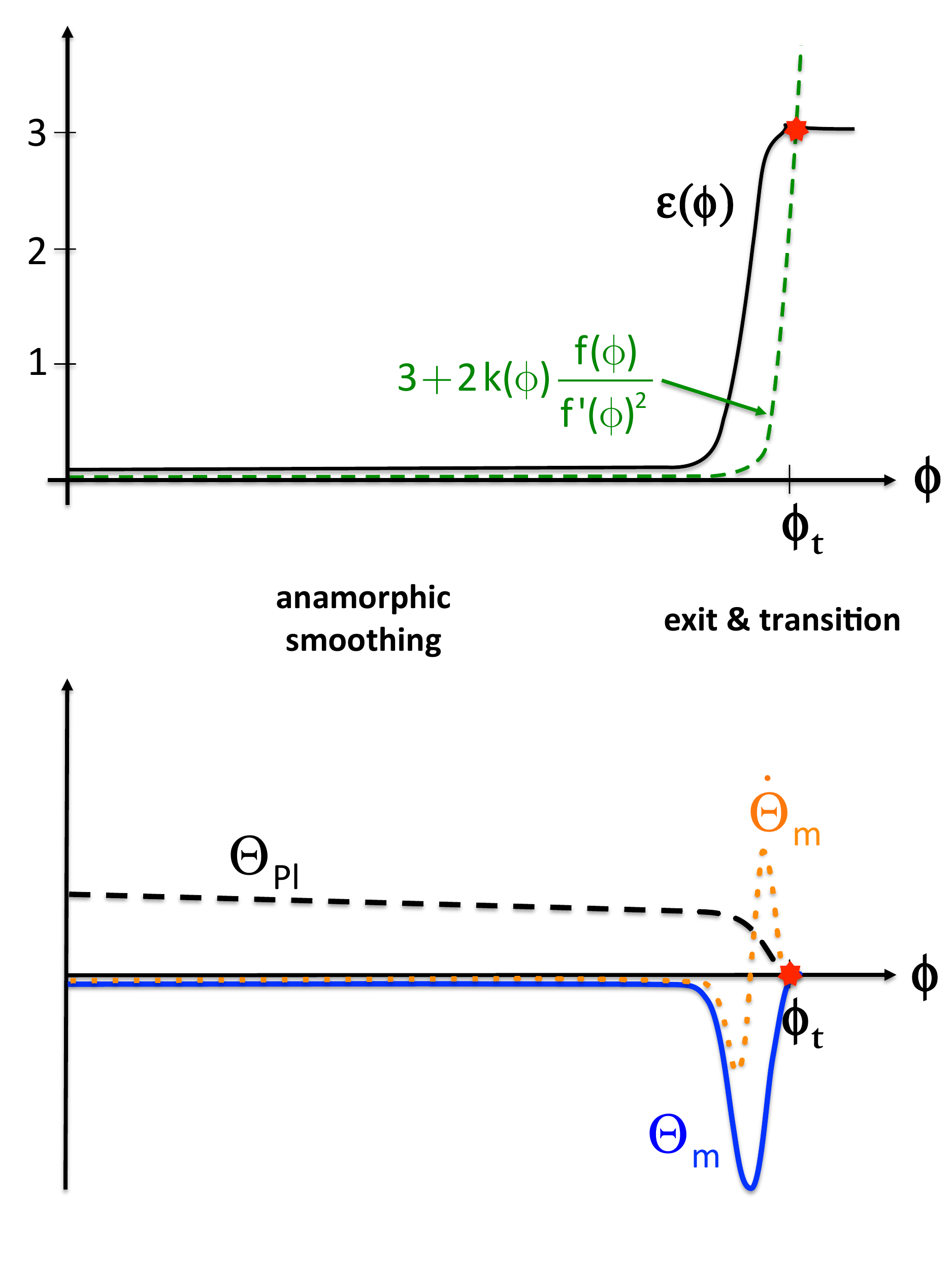}
\caption{ 
Plot of the equation-of-state parameter, $\epsilon$ (continuous black line in upper figure) and the frame-invariants, $\Theta_{\rm Pl}$ (dashed black line in lower figure) and $\Theta_m$ (continuous blue line in lower figure) as a function of $\phi \equiv \ln(A \chi)/A $ for the anamorphic Lagrangian~(\ref{Genaction}) with $f (\chi)= \xi \chi^2/(1+ \beta^2 \chi^2), k(\chi) = -1 + 2\gamma^2\chi^2/(1+ \gamma^2 \chi^2)$ and $V_J(\chi) = \lambda \chi^q \exp(-\alpha\chi)$ and the parameter values $\xi = 0.167, q = 3.99, \alpha = 1, \beta = 0.01$, and $\gamma = 0.05$. The dotted orange line in the lower figure is the (Jordan-frame) time-derivative of $\Theta_m$. The dashed green line in the upper figure is the function defined in Eq.~(\ref{contrJF2}), $3 + 2kf/f'^2$.   When $3 + 2kf/f'^2 = \epsilon$ the transition from (ordinary) contraction to (decelerated) expansion occurs; $\phi_t$ ($ \lesssim M_{\rm Pl}^0 \equiv 1$) is the field value at the transition.  The $y$-axis of the lower figure has arbitrary units. 
}
\label{fig:2}
\end{figure}

In Fig.~\ref{fig:1},  we have drawn a typical anamorphic potential with gravitational and kinetic couplings.   In Fig.~\ref{fig:2}, we have plotted the corresponding equation-of state parameter $\epsilon$ and the frame-invariants, $\Theta_m$ and $\Theta_{\rm Pl}$ (all as a function of the anamorphic field $\phi$) up to the transition to reheating and the hot big-bang phase.  Although these are sketches, all the properties described by these figures and and in the discussion below have been confirmed by explicitly solving specific examples, such as the one detailed in the caption to Fig.~\ref{fig:2}.  The solutions were found numerically by solving the differential equation Eq.~\eqref{epsN} for $\epsilon$ as a function $\phi$, and then 
substituting the solution into the following relations:
\begin{eqnarray}
 \Theta_{\rm Pl}(\phi) &= & \sqrt{\frac{V_J}{f^2 (3-\epsilon)} }, 
 \\
  \Theta_m(\phi)  & = & \Theta_{\rm Pl}(\phi) - \frac{1}{2} \frac{f,_{\phi}}{\, f} \frac{\dot{\phi}_J}{\sqrt{f}},  
  \\
   \dot{\phi}_J(\phi)  & =  & \Theta_{\rm Pl} \sqrt{\frac{2\epsilon}{K}\, f} 
.
   \end{eqnarray}
   The first relation follows from rewriting the Friedmann equations re-expressed in terms of the invariant $\Theta_{\rm Pl}$;  the second and third expressions follow from the definitions of $\Theta_m$, $\Theta_{\rm Pl}$ and $\epsilon$.  
   
\noindent   
The key features  of the simple scenario are as follows:

{\it Anamorphic smoothing:} 
The cosmological evolution in the simple scenario starts with the anamorphic smoothing phase at small field values (on the left hand side of  the barrier in Fig.~\ref{fig:1} where $\phi \ll \phi_t  \lesssim M_{\rm Pl}^0 \equiv 1$) and the potential energy density being very low.  In this phase,  $\epsilon$ is  less than one and slowly-varying, as given by Eq.~\eqref{SlowRoll1}.   Also, $\Theta_m <0$ (see lower panel of Fig.~\ref{fig:2}), corresponding to a universe that is contracting.
Due to the wrong-sign of the kinetic term $k$ the anamorphic field, $\phi$ moves uphill along the potential energy curve $V_J$ .  The gravitational coupling $f$ grows monotonically and so  the effective gravitational constant becomes weaker, contributing to the anamorphic smoothing process (recall Eq.~\ref{FriedmannEq1}).  $\Theta_{\rm Pl}$ is positive and very slowly decreasing, ensuring that the curvature perturbations are (nearly) scale-invariant with a small red tilt. The amount of anamorphic smoothing is given by the logarithmic change in $\alpha_{\rm Pl}$, $\ln (\alpha_0/\alpha_{\rm Pl})$, where $\alpha_0$ is the value of $\alpha_{\rm Pl}$ at the end of the smoothing phase, as in Eq.~\eqref{N}.  Quantum fluctuations generated during this period result in the scalar spectral tilt $n_S$ and  tensor-to-scalar ratio $r$  computed  in Sec.~\ref{subsec:simplePotential}, where the result (Case I or Case II) depends on the choice of parameters for $V_J$, $f$ and $k$.
 
{\it Exit and Transition:} 
The anamorphic smoothing phase comes to an end when the kinetic energy becomes comparable to the field's potential energy,  i.e., 
$\epsilon \simeq 1$.  Although the anamorphic phase ends before  $\phi$ reaches the maximum of the potential, $\phi$ continues to grow because the universe is still contracting (until  $\phi  =\phi_t$) and, hence, the kinetic energy of $\phi$ continues to blue shift. The value of  $\epsilon$ increases over the next interval because  of a combination of factors:  the potential   $V_J$ has a maximum, as depicted in Figure~\ref{fig:1}; the gravitational coupling $f$ increases with $\phi$;  and the kinetic coupling $k$ is  nearly constant $\sim -1$ (for $\phi<\phi_t$).  

$\dot{\Theta}_m$ reaches its most negative value and begins to increase after $\epsilon$ reaches $\epsilon \sim 1$ and the anamorphic phase ends.   Over the next period, because $\dot{\Theta}_m<0$,  $\Theta_m$ becomes more and more negative, but now at a slower rate than before due to the fact that $\dot{\Theta}_m$ is increasing towards zero.     Next, $\dot{\Theta}_m$ grows to the point that it reaches zero and, hence, $\Theta_m\ll0$ reaches its most negative value. As $\dot{\Theta}_m$  passes through zero and becomes positive, the field is still climbing uphill but ${\Theta}_m$ is becoming less negative. When $\dot{\Theta}_m$ hits its maximum, $\phi$ reaches the top of the potential and begins to roll downhill towards the minimum of the potential $V_J$.  At last, when the  kinetic coupling, $k$, passes through zero and changes from negative to positive at $\phi = \phi_t$, $\Theta_m$ also passes through zero and becomes positive.  This is  the moment we call the $\Theta_m$-bounce.  (In the  model proposed by Li  \cite{Li:2014qwa},  $k$ remains negative throughout, including the hot expanding phase when the reduced Planck mass becomes fixed; an equivalent of the $\Theta_m$-bounce is achieved by introducing tuned modifications of the effective potential which potentially induces an unintended inflationary phase that leads to a multiverse.)

Notably, the $\Theta_m$-bounce can occur  without violating the null-energy condition (NEC), as can  be seen from Eqs.~(\ref{JFeqsofmotion}--\ref{Hdot0}).   
 The $\Theta_m$-bounce occurs at finite values of the effective Weyl-invariant scale factor $\alpha_m$ and does not require a $\Theta_{\rm Pl}$-bounce. After the $\Theta_m$-bounce, the field settles at the minimum of the potential where $\phi$ obtains a large mass; $f$ and $k$ become fixed such that  $M_{\rm Pl} \Theta_m = M_{\rm Pl} \Theta_{\rm Pl} = H_E$ in agreement with Einstein gravity and standard big-bang evolution.
 
{\it No multiverse:} As discussed in Sec.~\ref{sec:secnomult}, the inflationary multiverse problem arises due to  rare quantum
fluctuations that kick the inflaton field in a direction that delays the end of the inflationary smoothing phase.   Rare regions where this occurs continue to inflate and grow exponentially faster than regions that end inflation, so the rare regions  soon occupy the overwhelming majority of the volume of the universe.   This process repeats {\it ad infinitum}, leading to runaway `eternal inflation' and a multiverse.  In ekpyrotic models, the problem is avoided because the analogous rare fluctuations extend the duration of the ekpyrotic smoothing, but regions in the smoothing phase are {\it contracting} in this case.  The regions undergoing the rare fluctuations remain negligibly small compared to typical regions that complete the smoothing phase, bounce and expand, and so there is no runaway smoothing or multiverse. 

In the anamorphic scenario, smoothing also occurs during a contracting phase, but the situation is more subtle than the ekpyrotic case.  For some rare fluctuations, the only effect is to extend the contracting phase, just as in the ekpyrotic model, and the same argument ensures that these regions cause no problems. However, some rare fluctuations can cause the evolution of the scalar field to reverse direction (e.g., towards the left in Fig.~\ref{fig:1}) compared to the normal classical evolution in the anamorphic phase.  In those regions,   $\Theta_m = \Theta_{\rm Pl} - (f,_{\phi}/2f) (\dot{\phi}/M_{\rm Pl})$  flips sign from negative to positive, corresponding to expansion.  If the expansion were inflationary and sustained for many $e$-folds, this could potentially cause runaway and produce a multiverse.   However, it is straightforward to show this does not occur.  If a quantum fluctuation gives the scalar field a strong kick in the wrong direction and the kinetic energy density is large compared to the potential energy density, there is expansion but no inflation.  Analytic estimates (and our numerical solutions) show that the expansion lasts for only a fraction of an $e$-fold before the field picks up enough speed for $\Theta_m$  to flip sign from positive to negative, leading to a long period of contraction.   Because the expansion caused by the rare fluctuation does not even last for an $e$-fold, there is no chance of runaway inflation, and no multiverse problem arises.

\subsection{Cyclic anamorphosis}
\label{sec:secCyclic}

In the preceding section we presented a one-time cosmological scenario where we started with the anamorphic smoothing phase, transited to standard big-bang evolution via a $\Theta_m$-bounce and ended with the current phase of dark-energy domination.  
In this section, we briefly outline an example of how the anamorphic scenario can be made cyclic, i.e., how a dark-energy phase  like  the present epoch can be connected to an anamorphic smoothing phase that seeds initial conditions for the next cycle of evolution; for a detailed description see our analysis in  \cite{anamorphic2}. A cyclic scenario has the advantage of avoiding initial condition problems related to assuming any sort of beginning in time. 
\begin{figure}[tb]
\begin{center}
\includegraphics[width=13.75cm]{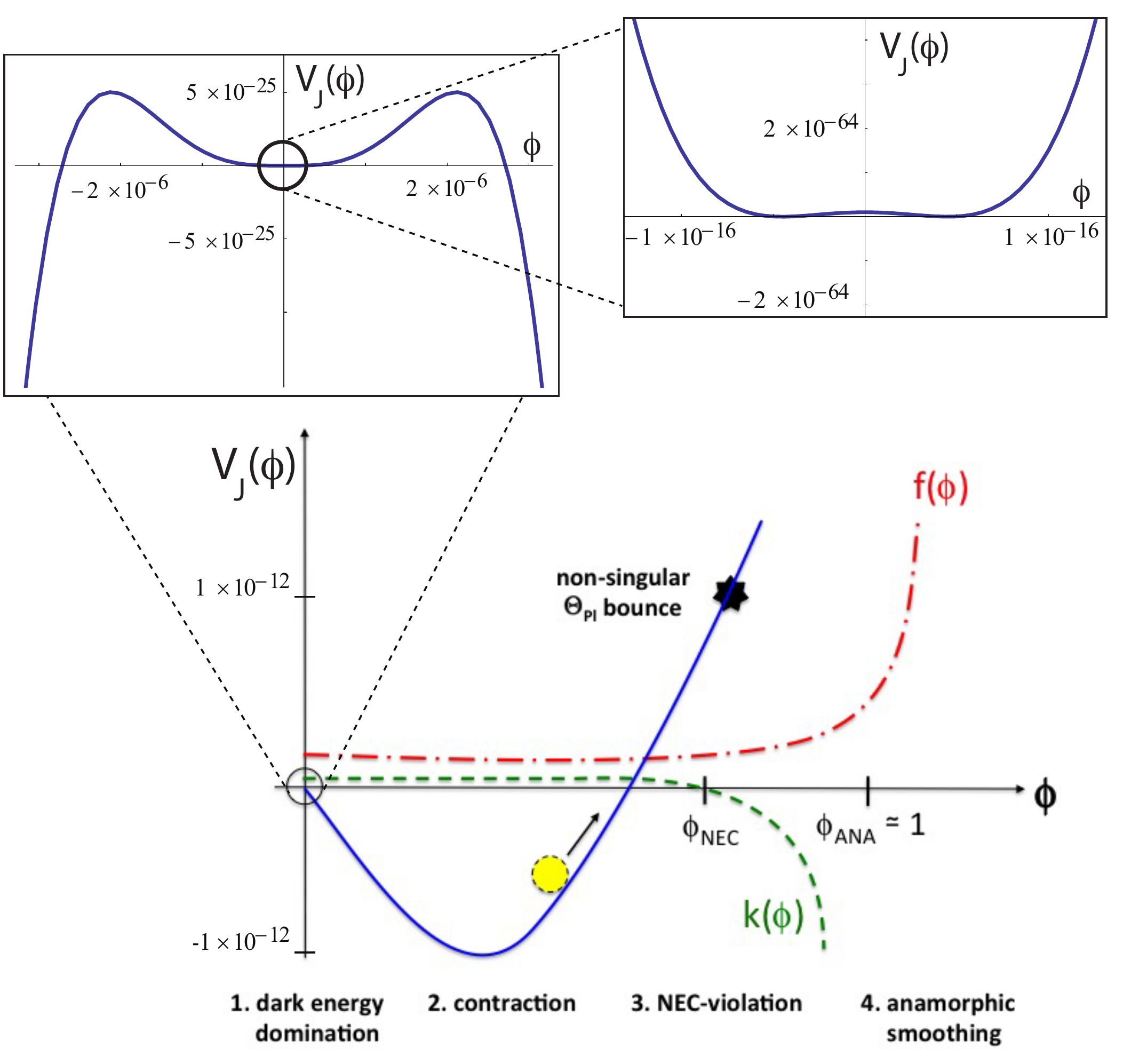}
\caption{Cyclic anamorphosis. 
A sequence of expanded views of the (Jordan-frame) potential $V_J$ (continuous blue line)  versus the anamorphic field $\phi$ in the cyclic anamorphic scenario with the middle inset showing the energy barrier and the final inset showing the current metastable vacuum. Both the anamorphic field $\phi$ and its potential $V_J$ are given in Planck units. We superposed curves to represent the gravitational coupling, $f$ (dotted-dashed red line) and the kinetic coupling, $k$ (dashed green line) that both are dimensionless functions of $\phi$. For small field values ($\phi\ll \phi_{\rm ANA}$) and, in particular, during the current cosmological epoch when the anamorphic field is trapped in the metastable vacuum, $f= 1$ and $k=1$, in agreement with observations; the plots of $f$ and $k$ have been shifted apart for the purpose of illustration.
}
\label{fig:cyclic}
\end{center}
\end{figure}
The key stages of the cyclic scenario are illustrated in Fig.~\ref{fig:cyclic}: 

{\it 1. Dark-energy domination} ({\it the present phase,} $\phi = \phi_0 \simeq 10^{-16} M_{\rm Pl}^0$):
For the purposes of illustration, we shall take the anamorphic field to be the Higgs field, where we will assume that the current vacuum ($\phi=\phi_0$) is metastable and $\phi_0$ has the empirically measured value.     In the current cosmological phase, the anamorphic field is settled in a local, metastable vacuum state of the Higgs-like potential $V_J(\phi)$, separated by a barrier of $\sim (10^{10-12}\, \mathrm{GeV})^4$ from the true, negative energy-density vacuum. 
The potential energy density $V_J(\phi_0)$ is the dominant energy component and it has the observed value of the current vacuum or dark-energy density $\sim (10^{-12}\, \mathrm{GeV})^4$; the gravitational coupling is $f(\phi_0) \equiv 1$ (in agreement with tests of general relativity today) and the kinetic coupling is $k \equiv 1 $ (in accordance with the standard Higgs model).
Due to the positive vacuum energy density of the metastable vacuum, the cosmological background expands at an accelerated rate.

2. {\it Contraction} ($10^{-6} M_{\rm Pl}^0 \lesssim \phi \lesssim \phi_{\rm NEC}$): 
Eventually, the metastable vacuum decays and $\phi$ rolls or tunnels to the negative part of the potential. As the field rolls downhill to more negative potential energy density, $\Theta_m=\Theta_{\mathrm{Pl}}$ both undergo a reversal from positive to negative and the universe starts contracting.  During the contracting phase that follows, the blue shifting kinetic energy density rapidly dominates the total energy density such that $\epsilon \gtrsim 3$.
 Rolling downhill, the anamorphic field keeps picking up kinetic energy and, due to its large kinetic energy density ($\epsilon \gg 3$) at the bottom of the potential, $V_{\mathrm{min}}$, the field does not settle in the true, negative vacuum but continues to increase and roll uphill while the background keeps contracting.

3. {\it NEC-violation and non-singular $\Theta_{\mathrm{Pl}}$-bounce} ($\phi_{\mathrm{NEC}} \lesssim \phi \lesssim \phi_{\mathrm{ANA}} \simeq M_{\rm Pl}^0 $):
Rolling uphill, the field's kinetic energy grows at a slower rate  but continues to dominate ($\epsilon \sim 3$)  for a substantial period after $V_J(\phi)$ changes from negative to positive.  The depth of the potential $|V_{min}|$ can be chosen so that  $\epsilon \sim 3$  when $\phi = \phi_{\mathrm{NEC}}$ where $k(\phi)$ changes from positive to negative, as illustrated in Fig.~\ref{fig:cyclic}.   
The sign-switch of the kinetic coupling leads to a brief violation of the null-energy condition (NEC). As we show in \cite{anamorphic2}, it is possible to achieve a ghost-free and  stable NEC-violating phase
by assuming that the effective Lagrangian for $\phi_{\rm NEC} \lesssim \phi \lesssim \phi_{\rm ANA}$ is Galileon-like \cite{Creminelli:2006xe,Nicolis:2008in,Chow:2009fm,Creminelli:2012my,Elder:2013gya}  with higher order kinetic terms, $\sim (\partial \phi)^4, \sim \Box\phi(\partial \phi)^2$. 
During NEC violation, $H_E = M_{\mathrm{Pl}} \Theta_{\mathrm{Pl}}$ begins less than zero, increases steadily and eventually hits zero and switches from negative to positive, resulting in a $\Theta_{\rm Pl}$-bounce. 
At the time of the $\Theta_{\mathrm{Pl}}$-bounce, the gravitational coupling $f$  is beginning to increase, as depicted in Fig.~\ref{fig:cyclic}.
It is at this point in the evolution that $\Theta_m$ and $\Theta_{\mathrm{Pl}}$ first become distinguishable. 
In particular, $\Theta_m$ remains finite and negative during and after the $\Theta_{\mathrm{Pl}}$-bounce, while $\Theta_{\mathrm{Pl}}$ becomes positive.
With increasing $f$, the quantity (1/2)$K(\phi)(\dot{\phi}/M_{\rm Pl})^2$  as given in Eq.~\eqref{defK} becomes positive and the NEC-violating phase ends. 
At this point, the couplings to higher order kinetic terms (functions of $\phi$) can be chosen that the terms are negligible as $\phi$ increases further.

4. {\it Anamorphic smoothing phase} ($\phi_{\mathrm{ANA}} < \phi$):
The conditions are now precisely what is required for the anamorphic smoothing phase. That is, during the first three stages, the dark energy-dominated phase ends by tunneling to a state with negative potential energy density; a period of contraction begins that drives the field uphill in $V_J$ through the brief period of NEC-violation that flips the sign of $\Theta_{\rm Pl}$ to be positive and opposite that of $\Theta_m$; and the NEC-violating phase ends at  approximately the value of $V_J(\phi)$ needed to generate energy-density perturbations of the right amplitude, as given by Eq.~\eqref{amplitude}.   (Note that, the important quantity for obtaining the right perturbation amplitude is $V_J/f^2$, which explains why it is possible in the simple scenario to have the anamorphic phase occur at small field values where $V_J(\phi)$ and $f$ are small, whereas here the field value, $V_J(\phi)$ and $f$ are all comparatively large.)  

Due to the wrong-sign of the kinetic term $k$ the anamorphic field $\phi$ continues to move uphill along the potential energy curve $V_J$ .  The gravitational coupling, $f$ grows monotonically and so  the effective gravitational constant becomes weaker, contributing to the anamorphic smoothing process.   
The Hubble-like parameters $\Theta_m<0$ and $\Theta_{\rm Pl}>0$ have opposite sign, though, both $\Theta_m^2$ and $\Theta_{\rm Pl}^2$ are very slowly shrinking and approaching zero during the anamorphic phase while the mass-variation index $q$ as defined in Eq.~\eqref{p} remains nearly constant. Notably, $\dot{\Theta}_m /M_{\rm Pl} > 0$ without violating the NEC.  The very slow decrease of $\Theta_m^2$ and $\Theta_{\rm Pl}^2$  ensures that the curvature perturbations are (nearly) scale-invariant with a small red tilt. The scalar and tensor spectra can be computed as described in Sec.~\ref{subsec:simplePotential}, where the result (Case I or Case II) depends on the choice of parameters for $V_J$, $f$ and $k$.

5. {\it  Exit, $\Theta_m$-bounce and hot big-bang evolution:}
The anamorphic smoothing phase ends when $\phi$ reaches infinity after finite time $\Delta t_J$ and $\Theta_m<0$ approaches zero.  We assume a bounce occurs, analogous to that in ekpyrotic models, in which $\dot{\phi}$ reverses sign and gets a small kick \cite{Bars:2013vba}.  $\Theta_m$ continues to increase, which means it switches from negative to positive resulting in  a  $\Theta_m$-bounce. Consequently, the universe starts expanding. No $\Theta_{\rm Pl}$-bounce occurs, though.  Rather  $\Theta_{\rm Pl}$  undergoes a reversal: it is positive before the $\Theta_m$-bounce, reaches zero at the $\Theta_m$-bounce and continues to decrease and becomes negative after the $\Theta_m$-bounce.   

On the way downhill, the gravitational coupling $f$ is shrinking and the kinetic coupling $k$ is growing, i.e., becoming less negative. The field eventually re-enters the range  of NEC-violation encountered on the way uphill. This occurs, when $(f,_{\phi}/f)^2$ has become small again relative to $|k/f|$. As before, $\dot{\Theta}_{\rm Pl}/M_{\rm Pl}>0$ leads to a $\Theta_{\rm Pl}$-bounce and $\Theta_{\rm Pl}$ becomes positive.
No $\Theta_m$-bounce occurs.  As $\phi$ continues to decrease and  $\dot{f} /M_{\rm Pl}$ vanishes, the invariants $\Theta_m, \Theta_{\rm Pl}$ (now both positive) as well as the frames become indistinguishable.  The field continues to roll down the potential energy curve, reaches the negative minimum and continues uphill. The small extra kick generated at the  $\Theta_m$-bounce enables the field to cross the tiny barrier and settle in the metastable vacuum state.  The Higgs field oscillates around its minimum converting its remaining kinetic energy to radiation, thereby reheating the universe.  After 14 billion years of expansion and cooling, the universe reaches a condition like the present, dominated by the small, positive false vacuum energy density of the Higgs field, and the cycle returns to step 1.

\section{Discussion}
\label{sec:secDiscussion}

The goal of this paper has been to introduce a new  approach for smoothing and flattening the early universe and generating a nearly scale-invariant spectrum of perturbations utilizing a single scalar field that couples non-minimally to gravity.  Through the invariants $\Theta_m$ and $\Theta_{\rm Pl}$, we have shown that we can classify different cosmological smoothing mechanisms and  specify how  the  anamorphic approach differs from previously known inflationary and ekpyrotic approaches (which may be formulated with or without non-minimal coupling).  It is clear from this classification that the anamorphic mechanism is a meld of the earlier ideas, aiming to employ the best features of both.

During the smoothing phase, the conditions on the equation of state $\epsilon$ and $\Theta_{\rm Pl}$ are reminiscent of inflation.  Much of the intuition and experience gained in constructing inflationary models can be usefully applied in constructing models of the anamorphic smoothing phase.  However, the differences are also important.  The universe is physically contracting in the anamorphic phase, as quantified by $\Theta_m$, which is key to avoiding the multiverse problem of inflationary models.  Also, there are enormous differences in  the conditions for embedding the smoothing phase in a complete cosmological model.  Inflation relies on a big bang to precede inflation and leads to eternal inflation.  To utilize the anamorphic mechanism, the big bang must be replaced by a $\Theta_m$-bounce, and the anamorphic phase must precede that bounce.  

In this respect, anamorphic cosmology is reminiscent of ekpyrotic cosmology.   For example, as we illustrated in one of our constructions, replacing the bang with a bounce opens the possibility of embedding the anamorphic mechanism in a geodesically complete cyclic theory of the universe that resolves both the initial conditions and multiverse problems.  (Notably, our construction includes a novel Galileon-like mechanism for producing a non-singular $\Theta_{\rm Pl}$-bounce that evades gradient instabilities and ghosts, as detailed in Ref.~\cite{anamorphic2}.) 

The specific examples presented in this paper were constructed to illustrate certain design principles.  In future papers, we will present examples that are more efficient and aesthetic.  We will also show how to generalize the anamorphic mechanism to more general types of scalar-tensor and other modified gravity theories.

\acknowledgments

We thank D. Eisenstein, M. Johnson, J. Khoury, E. Komatsu, J.-L. Lehners,  S. Perlmutter and N. Turok for helpful discussions. This research was partially supported by the U.S. Department of Energy under grant number DE-FG02-91ER40671 (PJS). 

\appendix

\section{Cosmological background in Jordan and Einstein frames}
\label{appendix:background}

We present the background equations of motion for the anamorphic action in Eq.~(\ref{Genaction}) in both Jordan  and Einstein  frames.

\subsection{Jordan-frame representation}
\label{sec:secbackgrounds1}

The Jordan frame with  $M_{\textrm{Pl}} \Theta_m = H_J \neq M_{\textrm{Pl}} \Theta_{\rm Pl}$  has the property that the dust particle mass $m_J$ is independent of $\phi$.  Consequently, this is the frame in which physical rulers (set by the particle's Compton wavelength, say)  have fixed length.  Varying the action with respect to the metric and the field leads to the (Jordan-frame) equations of motion
\begin{eqnarray}\label{JFeqsofmotion}
3f(\phi) \Big( H_J^2 + \frac{\kappa}{a_J^2}\Big) - \frac{\sigma^2}{f(\phi)\,a_J^6}&=& \frac{1}{2}k(\phi)\dot{\phi}_J^2 + V_J(\phi) - 3H_Jf,_{\phi}\dot{\phi}_J + \rho_m^J,\\
\label{Hdot0}
f(\phi)\Big(\dot{H}_J - \frac{\kappa}{a_J^2}\Big) + \frac{\sigma^2}{f(\phi)\,a_J^6} &=& -\frac{1}{2}k(\phi)\dot{\phi}_J^2 + \frac{1}{2}H_J{f,_{\phi}}\dot{\phi}_J - \frac{1}{2}\left(f,_{\phi\phi}\dot{\phi}_J^2+f,_{\phi}\ddot{\phi}_J\right) \nonumber\\
&&- \frac{1}{2} \left(\rho_m^J +  p_m^J \right),\\
\label{fieldeq0}
k(\phi) \big( \ddot{\phi}_J + 3H_J\dot{\phi}_J \big) + V_J,_{\phi} &=&  3f,_{\phi}\Big( \dot{H}_J + 2H_J^2 + \frac{\kappa}{a_J^2} + \frac{\sigma^2}{3f^2\,a_J^6} \Big) - \frac{1}{2} k,_{\phi}\dot{\phi}_J^2,
\\ 
\label{anisotropy}
\frac{1}{2}\sum_{i}\dot{\beta}_i^2 &=& \frac{\sigma^2}{f^2(\phi)a_J^6},
\end{eqnarray}
where we define  $f(\phi) \equiv (M_{\rm Pl} (\phi)/M_{\rm Pl}^0)^2$, where $M_{\rm Pl}^0$ is the value of the reduced Planck mass today. 
The dot denotes differentiation with respect to the physical (Jordan frame) time, $t_J$,  that runs from large negative to small negative values during the anamorphic phase; although the value of $\phi$ is frame invariant, time derivatives such as $\dot{\phi}_J$ are not so we add the $J$ subscript to be clear that the derivative refers to Jordan frame time.   $\rho_m^J$ is the (Jordan-frame) matter energy density and $p_m^J$ is the (Jordan-frame) matter pressure.
The spatial curvature is $\kappa = (+1,\, 0,\, -1)$ and a homogeneous and anisotropic Kasner-like metric is assumed where 
\begin{equation}\label{kasner}
ds_J^2 = -dt_J^2 + a_J^2(t)\sum_i \exp(2\beta_i)dx_i^2  \quad \text{   with  } \quad \sum_i\beta_i(t) = 0.
\end{equation}
In the field equations, Eq.~(\ref{JFeqsofmotion}--\ref{anisotropy}), the anisotropy is parameterized by 
$\sigma^2$ where 
\begin{equation}\label{sigma} 
\sigma^2 = \frac{1}{2}\sum_i c_i^2, \quad \beta_i = \frac{c_i}{fa_J^3},\quad c_i \equiv \text{constant}. 
\end{equation}

Smoothing and flattening the universe and generating nearly-scale invariant perturbations during the anamorphic phase  and transitioning afterwards to a hot expanding phase imposes constraints on $f(\phi)$, $k(\phi)$ and $V_J(\phi)$.  During the anamorphic phase, the invariant mass ratio $m/M_{\rm Pl}$ must shrink or, equivalently, the non-minimal coupling $f(\phi)$ must grow significantly to suppress the anisotropy.  Then, during the hot expanding phase that follows, including the present epoch, $f(\phi)$ must approach unity so that  $M_{\rm Pl}^2 \to (M_{\rm Pl}^0 )^2\equiv 1$.
 These conditions ensure that there are no observable deviations from Einstein gravity today or at observable red shifts.   

 In the hot expanding phase where $f(\phi)$ is constant, $k(\phi)$ must be positive (the standard sign in simple quantum field theories) so that the kinetic energy density is non-negative and  there is no ghost-field.   However, during the anamorphic phase when $f(\phi)$ is growing, $k(\phi)$ must be negative in order to obtain the equation-of-state, $\epsilon$, required for smoothing and flattening, as has been shown in Sec.~\ref{sec:ScalarTensor}.   Hence, the kinetic coupling $k(\phi) $ must change from negative during the anamorphic phase to positive during the subsequent hot expanding phase that follows. Because  $k(\phi)$ varies from negative to positive during the scenario, it cannot be eliminated by a field re-definition in the Jordan-frame action.  Smoothing also requires that the potential  energy density  $V_J(\phi)$ be positive and increasing during the anamorphic phase. This is achieved as a result of the negative (wrong-sign) kinetic coupling $k(\phi)$ during the anamorphic phase, which causes the solution for $\phi$ to run uphill along its potential $V_J(\phi)$.  (By contrast, the analogous field during an ekpyrotic contracting phase has normal-sign kinetic energy density and runs downhill along a potential that is negative and decreasing.)  

\subsection{Einstein-frame representation}

The Jordan frame is natural  for describing the smoothing and flattening of the universe and the different stages of expansion and contraction before and after the smoothing phase because rulers have fixed lengths in this frame.     However, for some purposes, it is useful to consider the action in the Einstein frame where  $M_{\mathrm{Pl}}\Theta_{\rm Pl}= H_E \neq M_{\mathrm{Pl}} \Theta_m$, especially when  comparing gravitational effects, such as wavelengths of  curvature perturbation modes compared to the Hubble radius. 
Under a Weyl transformation of the Jordan-frame metric, the Einstein-frame metric is
\begin{equation}
\label{EFmetric}
g_{\mu\nu}^E = f(\phi)g_{\mu\nu}^J. 
\end{equation}
Introducing the field $\Phi$ given by
\begin{equation}\label{defPhi}
\frac{d\Phi}{d\phi} = \sqrt{K(\phi)},
\end{equation}
where
\begin{equation}\label{defK}
K(\phi) \equiv \frac{\frac{3}{2}( f,_{\phi})^2 + k(\phi)f(\phi)}{f^2(\phi)}, 
\end{equation}
the gravitational and field sector of the action can always be rewritten in terms of minimally-coupled Einstein gravity,
\begin{equation}
S_E = \int d^4x \sqrt{-g_E}\left(\frac{1}{2}R_E - \frac{1}{2}(\partial_{\mu}\Phi)^2 - V_E(\Phi)  \right),
\end{equation}
where
\begin{equation}
V_E(\Phi) \equiv V_J(\phi)/f^2(\phi).
\end{equation}
During the anamorphic smoothing phase, both $V_E$ and $V_J$ are positive.  

 $K(\phi)$ must be positive definite and continuously differentiable in order for the field to have the right-sign kinetic term in the Einstein frame. Hence, 
\begin{equation}\label{noghost}
K(\phi) > 0
\end{equation}
is the {\it no-ghost condition}. From the expression for $K(\phi)$ in Eq.~(\ref{defK}), we see that $k(\phi)$ can be negative  and the theory can remain ghost-free if $(f_{,\phi})^2/f$ is sufficiently large.  

\subsection{Matter contribution and contraction}
\label{sec:sec2pt3}

The dust contribution to the action in the two frames is such that the action of a single test particle is given as in Eq.~\eqref{Dustaction}, 
\begin{equation}
\label{JFmass}
S_{d} = \int  m_J ds_J = \int  m_J/\sqrt{f(\phi)}\, ds_E.
\end{equation}
The dust-particle mass in Einstein-frame representation is given by 
\begin{equation}\label{EFmass1}
m_E = m_J/\sqrt{f(\phi)}. 
\end{equation}
In the anamorphic models considered in this paper,  $m_J$ is $\phi$-independent and, hence, $m_E$ is $\phi$-dependent.    
Because $m_J$ is fixed independent of $\phi$, having  $ H_J<0 $ (or, equivalently, $\Theta_m <0$), as occurs during the anamorphic phase, means that the universe is contracting compared to physical rulers made from matter.   However,  as we have noted, the Hubble parameter in other frames may have a different sign.   In particular, $H_E$ (or, equivalently, $\Theta_{\rm Pl}$) is positive in the anamorphic phase. 

\section{Conditions for successful anamorphosis}
\label{appendix:Conditions}

We analyze  the conditions that  $f(\phi)\equiv (M_{\rm Pl}/M_{\rm Pl}^0)^2$, $k(\phi)$ and $ V_J(\phi)$  in the action~\eqref{Genaction} and  that the equation of state $\epsilon(\phi)$  must satisfy in order to smooth and flatten the cosmological background and to have squeezed quantum fluctuations during the anamorphic contracting phase  ($\Theta_m < 0<\Theta_{\rm Pl}$).

{\it Constraint 1: Contraction.}
The invariant anamorphic energy density is given by the sum of the kinetic and potential energy density of the anamorphic field $\phi$ and the invariant pressure is given by the difference: 
\begin{eqnarray}
\frac{\rho_A}{M_{\rm Pl}^4} &\equiv& \frac{1}{2} K(\phi) \left( \frac{\dot{\phi}}{M_{\rm Pl}^0 M_{\rm Pl}} \right)^2 + \frac{V_J(\phi)/f^2(\phi)}{(M_{\rm Pl}^0)^4} ;
\\
\frac{p_A}{M_{\rm Pl}^4} &\equiv& \frac{1}{2} K(\phi) \left( \frac{\dot{\phi}}{M_{\rm Pl}^0 M_{\rm Pl}} \right)^2 - \frac{V_J(\phi)/f^2(\phi)}{(M_{\rm Pl}^0)^4},
\end{eqnarray}
where for pedagogical reasons we have kept factors of $M_{\rm Pl}^0$ exposed.
Since the energy density of the anamorphic scalar dominates the total energy density during the smoothing phase, the conventional equation-of-state parameter, $\epsilon \equiv (3/2)(1+w)$ with $w\equiv p/\rho$, 
can be re-expressed as
\begin{equation}
\label{epsilon}
\epsilon \equiv \frac{1}{2} \frac{K(\phi)}{\Theta_{\rm Pl}^2} \left( \frac{\dot{\phi}}{M_{\rm Pl}^0 M_{\rm Pl} } \right)^2,
\end{equation}
so that the anamorphic condition, $\Theta_m<0$ (contraction relative to physical rulers) and $\Theta_{\rm Pl} > 0$ (expansion relative to the Planck scale), becomes
\begin{eqnarray}
\label{HJF}
\Theta_m  
= 
\Theta_{\rm Pl} + M_{\rm Pl}^{-1} \left( \frac{\dot{m}}{m} - \frac{ \dot{M}_{\rm Pl} }{ M_{\rm Pl} }  \right)
= \left( \sqrt{ \frac{ K(\phi)}{2\epsilon}} - \frac{1}{2} \frac{f,_{\phi}}{f} \right) \frac{\dot{\phi}_J}{M_{\rm Pl}}
< 0. 
\end{eqnarray}
Without loss of generality, we will consider the case when $\phi$ is increasing during the anamorphic phase,  $\dot{\phi}/M_{\rm Pl} > 0$.
Since $K(\phi)$ and $\epsilon$ are positive definite, the only way $\Theta_m$ can be less than zero with $\dot{\phi}/M_{\rm Pl} > 0$ is if
\begin{equation}
\label{f_prime}
f,_{\phi}>0.
\end{equation}
(The constraint is $f,_{\phi}<0$ if $\dot{\phi}/M_{\rm Pl} < 0$.)
Then, substituting the definition of $K(\phi)$ from Eq.~(\ref{defK}) into Eq.~(\ref{HJF}), the condition that $\Theta_m<0$ reduces to
\begin{equation}
\label{contrJF2}
3+ 2k(\phi) \frac{f(\phi)}{f,_{\phi}^2} < \epsilon.
\end{equation}
It is straightforward to show that the constraint in 
Eq.~(\ref{contrJF2}) is the same independent of the sign of $\dot{\phi}/M_{\rm Pl}$.  

%%%%%%%%%%%%%%%%%%%%%

{\it Constraint 2: Smoothing of the Cosmological Background.}
Smoothing and flattening during contraction requires that the scalar field energy density dominates over all other contributions  for an extended period of 60 or more $e$-folds.  
We have shown that the necessary condition for smoothing and flattening is that $\epsilon_m>1-q$ or, equivalently, $\epsilon_{\rm Pl}\equiv \epsilon <1$ (see Eqs.~(\ref{smcon}) and~(\ref{ePLc})).
That is, smoothing and flattening imposes an upper-bound on $\epsilon$ which complements the lower bound constraint imposed by contraction, Eq.~(\ref{contrJF2}).
  
%%%%%%%%%%%%%%%%%%%%%%%%%

{\it Constraint 3: Squeezing of quantum curvature perturbations.} The squeezing constraint in Eq.~\eqref{sq1} can be rewritten as 
\begin{equation}
\frac{d \alpha_{\rm Pl}\Theta_{\rm Pl}}{d\,t}M_{\rm Pl}^{-1} = \alpha_{\rm Pl}\Theta_{\rm Pl}^2 \left( 1 + \frac{d\ln \Theta_{\rm Pl}}{d\ln \alpha_{\rm Pl}} \right) > 0.
\end{equation}
Since $\epsilon \equiv d\ln \Theta_{\rm Pl}/d\ln \alpha_{\rm Pl}$, the squeezing constraint is equivalent to the smoothing condition $\epsilon<1$. 

\hspace{1pt}

\noindent
Combining all three constraints with the no-ghost condition ($K>0$), we  have
\begin{equation}
\label{comb2}
0<3+ 2k(\phi) \frac{f(\phi)}{f,_{\phi}^2} < \epsilon<1.
\end{equation}

\bibliographystyle{plain}
\bibliography{anamorphicU}
\end{document}